\documentclass[a4paper,11pt]{article}
\pdfoutput=1 

\usepackage[utf8]{inputenc}
\usepackage{jheppub}
\usepackage{amsmath}
\usepackage{amsfonts}
\usepackage{amssymb}
\usepackage{mathtools}
\usepackage{amsthm}
\usepackage{dsfont}
\usepackage{hyperref}
\usepackage[capitalise,sort]{cleveref}
\usepackage{pgfplots}
\usepackage{tabularx}
\usepackage{environ}
\usepackage[inline]{enumitem}
\usepackage{comment}
\usepackage{tikz}
\usetikzlibrary{decorations.pathmorphing,calc,fit,matrix,cd,shapes,arrows,positioning, intersections,chains,positioning,arrows.meta,hobby,patterns,shadings,shadows}
\usepackage{blkarray}
\usepackage{shadow}
\usepackage{fancybox}
\usepackage{bm}
\usepackage{lscape}
\usepackage{verbatim}
\usepackage[normalem]{ulem}
\usepackage{subcaption}
\usepackage{mdframed}
\usepackage{orcidlink}

\pgfplotsset{
        compat=1.9,
        compat/bar nodes=1.8,
    }

\makeatletter
\def\@xfootnote[#1]{%
	\protected@xdef\@thefnmark{#1}%
	\@footnotemark\@footnotetext}
\makeatother

\definecolor{prhigh}{HTML}{ff0000}
\definecolor{sechigh}{HTML}{e0fbfc}
\definecolor{prcolor}{HTML}{1d3557}
\definecolor{seccolor}{HTML}{457b9d}
\definecolor{tercolor}{HTML}{98c1d9}
\definecolor{blueplot}{HTML}{58468e}

\newcommand\bea{\begin{eqnarray}}
\newcommand\eea{\end{eqnarray}}

\theoremstyle{plain}

\theoremstyle{definition}

\newtheorem{conjecture}{Conjecture}
\newtheorem*{conjecture*}{Conjecture}
\newtheorem{remark}{Remark}
\newtheorem{remark*}{Remark}



\DeclareMathOperator{\U}{U}

\newcommand{\ID}{\mathds{1}}

\newcommand{\coma}{\, , \quad}
\newcommand{\fstop}{\, .}

\newcommand{\AdS}{\text{AdS}}

\newcommand{\PlD}[1]{\text{\tiny Pl,\,#1}}

\newcommand{\UoD}[1]{\text{\tiny U(1),\,#1}}
\newcommand{\ttiny}[1]{\text{\tiny #1}}
\newcommand{\WGCD}[1]{\text{\tiny WGC,\,#1}}
\newcommand{\AdSD}[1]{\text{\tiny AdS,\,#1}}
\newcommand{\MinkD}[1]{\text{\tiny Mink,\,#1}}

\renewcommand{\epsilon}{\varepsilon}

\makeatletter
\newsavebox{\measure@tikzpicture}
\NewEnviron{scaletikzpicturetowidth}[1]{%
  \def\tikz@width{#1}%
  \begin{lrbox}{\measure@tikzpicture}%
  \BODY
  \end{lrbox}%
  \pgfmathparse{#1/\wd\measure@tikzpicture}%
  \BODY
}
\makeatother

\makeatletter
\newcommand{\inlineitem}[1][]{%
\ifnum\enit@type=\tw@
    {\descriptionlabel{#1}}
  \hspace{0pt}%
\else
  \ifnum\enit@type=\z@
      \hspace{-15pt} \refstepcounter{\@listctr}\fi
    \quad\@itemlabel\hspace{0pt}%
\fi}
\makeatother
\DeclareMathAlphabet{\mathdutchcal}{U}{dutchcal}{m}{n}

\tikzset{
    partial ellipse/.style args={#1:#2:#3}{
        insert path={+ (#1:#3) arc (#1:#2:#3)}
    }
}

\tikzset{cross/.style={cross out, draw=black, fill=none, minimum size=2*(#1-\pgflinewidth), inner sep=0pt, outer sep=0pt}, cross/.default={2pt}}

\tikzset{
	pics/torus/.style n args={3}{
		code = {
			\providecolor{pgffillcolor}{rgb}{1,1,1}
			\begin{scope}[
				yscale=cos(#3),
				outer torus/.style = {draw,line width/.expanded={\the\dimexpr2\pgflinewidth+#2*2},line join=round},
				inner torus/.style = {draw=pgffillcolor,line width={#2*2}}
				]
				\draw[outer torus] circle(#1);\draw[inner torus] circle(#1);
				\draw[outer torus] (180:#1) arc (180:360:#1);\draw[inner torus,line cap=round] (180:#1) arc (180:360:#1);
			\end{scope}
		}
	}
}

\tikzset{
	pics/hole/.style n args={2}{
		code = {
			\draw[fill=white] (0,0) arc(120:60:#1 and #2)  arc(-60:-120:#1 and #2);
            \draw (0,0) arc(-120:-130:#1 and #2) (#1,0) arc(-60:-50:#1 and #2);
		}
	}
}

\makeatletter
\newcommand*{\itemequation}[3][]{%
  \item
  \begingroup
    \refstepcounter{equation}%
    \ifx\\#1\\%
    \else  
      \label{#1}%
    \fi
    \sbox0{#2}%
    \sbox2{$\displaystyle#3\m@th$}%
    \sbox4{\@eqnnum}%
    \dimen@=.5\dimexpr\linewidth-\wd2\relax
    \ifcase
        \ifdim\wd0>\dimen@
          \z@
        \else
          \ifdim\wd4>\dimen@
            \z@
          \else 
            \@ne
          \fi 
        \fi
      \@latex@warning{Equation is too large}%
    \fi
    \noindent   
    \rlap{\copy0}%
    \rlap{\hbox to \linewidth{\hfill\copy2\hfill}}%
    \hbox to \linewidth{\hfill\copy4}%
    \hspace{0pt}
  \endgroup
  \ignorespaces 
}
\makeatother  

\hypersetup{
	pdftitle={Formulating the Weak Gravity Conjecture in AdS Space},    
	pdfauthor={\textcopyright\ Puxin Lin, Alessandro Mininno, Gary Shiu},     
	pdfsubject={hep-th},   
	pdfcreator={pdfLaTex},   
	pdfproducer={LaTex}, 
	pdfkeywords={},
	colorlinks=true
  ,urlcolor=blue
  ,anchorcolor=blue
  ,citecolor=blue
  ,filecolor=blue
  ,linkcolor=blue
  ,menucolor=blue
  ,linktocpage=true
}

\allowdisplaybreaks[1] 

\crefname{figure}{Figure}{Figures}
\crefname{table}{Table}{Tables}
\crefname{definition}{Definition}{Definitions}
\crefname{proposition}{Proposition}{Propositions}
\crefname{claim}{Claim}{Claims}
\crefname{conjecture}{Conjecture}{Conjectures}

\def\beq{\begin{equation}}
\def\eeq{\end{equation}}
\def\bc{\begin{cases}}
\def\ec{\end{cases}}
\def\bal{\begin{aligned}}
\def\eal{\end{aligned}}
\newcommand{\mc}{\mathcal}
\newcommand{\mf}{\mathfrak}

\arxivnumber{}
\title{Formulating the Weak Gravity Conjecture in AdS Space}
\author{Puxin Lin\,\orcidlink{0000-0003-0287-1662},}
\author{Alessandro Mininno\,\orcidlink{0000-0002-9593-0440},}
\author{Gary Shiu\,\orcidlink{0000-0003-1308-5202}}
\affiliation{Department of Physics, University of Wisconsin--Madison\\
1150 University Avenue, Madison, WI 53706, USA}

\emailAdd{plin73@wisc.edu}
\emailAdd{mininno@physics.wisc.edu}
\emailAdd{shiu@physics.wisc.edu}

\abstract{
We propose a version of the Weak Gravity Conjecture that applies to AdS spacetime. We find that the condition on the charge-to-mass ratio of a charged particle in AdS$_D$ spacetime is corrected compared to the one in Minkowski spacetime by contributions that depends on the AdS scale and the horizon radius of the extremal Reissner--Nordstr\"om black hole charged under the same gauge theory. It is maximized when we consider the largest possible extremal black hole in AdS. We motivate our proposal from the viewpoint of extremal black hole decay and show that the bound on the particle spectrum is given by the critical charge-to-mass ratio beyond which the Schwinger effect can take place. This quantum effect shares the same condition as requiring a particle to satisfy a repulsive force condition at the black hole horizon, so that the extremal black hole can decay without reabsorbing the particle. We discuss the relation of our proposed weak gravity bound with the near-horizon Breitenlohner--Freedman bound.
We also comment on the generalization in the case of multiple U$(1)$ gauge theories, providing evidence for a convex hull condition in AdS background. 
}

\begin{document}

\maketitle

\section{Introduction}
\label{sec:intro}

The Swampland program \cite{Vafa:2005ui} (see also \cite{Brennan:2017rbf,Palti:2019pca,vanBeest:2021lhn,Grana:2021zvf,Agmon:2022thq, Montero:2024qml} for reviews) seeks the universal features of Effective Field Theories (EFTs) coupled to gravity that have a UV completion --- those that are consistent are said to be in the Landscape and otherwise in the Swampland. One of the most important conjectures in the Swampland program is the Weak Gravity Conjecture (WGC) \cite{Arkani-Hamed:2006emk} (reviewed in \cite{Palti:2020mwc,Harlow:2022ich,Rudelius:2024mhq}), which puts a constraint on the spectrum of massive particles charged under an Abelian gauge symmetry. The main motivation for the WGC is to allow extremal Reissner--Nordstr\"om (RN) black holes to decay, postulating the existence of a charged massive particle with a charge-to-mass ratio larger than that of an extremal black hole (BH). Many versions of the WGC have since been proposed, scrutinized and refined. To date, the WGC in Minkowski space has passed all tests in the string theory setup.
It is therefore a likely criterion to discriminate between theories belonging to the Landscape and the Swampland.

Despite progress \cite{Cheung:2018cwt, Hamada:2018dde,Montero:2018fns}, a strict proof of the WGC is lacking. There is still room for further refinement of the WGC, as well as extension to other gravitational backgrounds. In this paper, we extend the formulation of the WGC to anti-de Sitter (AdS) spacetimes. Although there have been discussions of the WGC in AdS space \cite{Nakayama:2015hga,Montero:2016tif,Crisford:2017gsb,Yu:2018eqq,Cremonini:2019wdk,Horowitz:2019eum,McPeak:2020iwq,Aharony:2021mpc,Moitra:2023yyc,Ezroura:2024xba,Anand:2024wxa}, these generalizations lack a physical reasoning for why the proposed form should hold in AdS. In this paper, we motivate our proposal of the WGC conjecture in AdS with the idea of extremal black hole decay and support the conjecture with analysis of the charged decay process. Further evidence is provided by the consistency between the Schwinger effect of extremal Reissner--Nordstr\"om AdS (RN-AdS) black holes, the quantum effect, and the classical requirement that the produced charged particles be repelled by the black hole. The latter is reminiscent of the Repulsive Force Conjecture (RFC) \cite{Heidenreich:2019zkl}, stemming from the idea that the gravitational force should be weaker than the gauge force. The RFC proposes that in any EFT with gravity coupled to a $\U(1)$ gauge theory, there must exist a charged particle that is self-repulsive.

In the absence of massless scalar fields, the two conjectures have been shown to be equivalent, since in flat space, self-repulsiveness and superextremality become equivalent.\footnote{However, even in the presence of scalar fields, the towers of states satisfying the Sub-Lattice Weak Gravity Conjecture \cite{Heidenreich:2016aqi, Montero:2016tif} and the Tower Weak Gravity Conjecture \cite{Andriolo:2018lvp}
are also those satisfying the RFC \cite{Heidenreich:2019zkl,Lee:2018spm,Lee:2018urn,Lee:2019wij,Klaewer:2020lfg,Gendler:2020dfp,Cota:2022yjw,Cota:2022maf,FierroCota:2023bsp,Heidenreich:2024dmr}.} 

\subsubsection*{Summary of the Results}
In this work, we investigated the conditions under which an extremal RN-AdS black hole can decay. This study is based on the extension of the results in \cite{Lin:2024jug}, where the authors found the bound on the charge-to-mass ratio of a particle produced by Schwinger effect on the horizon of the RN-AdS black hole. We show that the same bound can be obtained by considering the Schwinger effect as an instability at the near-horizon $\AdS_2\times S^{D-2}$ geometry of the RN-AdS black hole, leading to charged particles that have an effective mass below the BF bound of the $\AdS_2$ space. We interpret this bound as the AdS version of the WGC in its original formulation, which states the requirement for an extremal charged black hole to decay. To confirm this statement, we interpret the WGC in its RFC formulation to find the conditions under which a particle experiences repulsion by the black hole so that it can stay separated from the black hole and not be reabsorbed. Unlike the flat space situation, however, the negative cosmological constant will always make the force between a particle and a black hole attractive when they are sufficiently far away from each other. However, by requiring the particle to feel repelled at the horizon of the black hole, the bound on the charge-to-mass ratio is the same as the one obtained via the two previous computations. This confirmed our interpretation that this condition represents the extension of the WGC in AdS. 

We, then, summarize our result in the following conjecture:
\begin{mdframed}[backgroundcolor=white, shadow=true, shadowsize=4pt,shadowcolor=seccolor,
roundcorner=6pt]
    \begin{conjecture}[AdS Weak Gravity Conjecture]\label{conj:WGCinAdS}
    Given any $\U(1)$ gauge field coupled to Einstein--Maxwell gravity, there must exist a particle with charge $\mf{q}$ and mass $\mf{m}$ such that
    \begin{equation}
       \frac{g_\UoD{D}\mf{q}}{\mf{m}}\ge \sqrt{\gamma}\frac{1}{M_\PlD{D}^{\frac{D-2}{2}}} \sqrt{\frac{1+\frac{(D-1)(D-2)}{(D-3)^2}\eta^2}{1+\frac{D-1}{D-3}\eta^2}}\coma \text{with } \eta = \frac{r_h}{\ell_\AdSD{D}}\coma
       \label{eq:chargetomassforconjecture}
    \end{equation}
    where $r_h$ is the horizon of an extremal RN-AdS black hole and $\ell_\AdSD{D}$ is the $\AdS_D$ scale length. The value $\gamma$ is the extremality factor for a RN black hole in flat space, i.e. $\gamma  = \frac{D-3}{D-2}$. For particles with a mass comparable to the AdS scale, we refer to Remark \ref{remark:BFbound}.
    Considering the largest possible RN-AdS black hole, i.e. $r_h\gg \ell_\AdSD{D}$, we require that there exists a particle of charge $\mf{q}$ and mass $\mf{m}$ such that 
    \begin{equation}\label{eq:chargetomassforconjecture-maximum}
       \frac{g_\UoD{D}\mf{q}}{\mf{m}}\ge \frac{1}{M_\PlD{D}^{\frac{D-2}{2}}}\fstop
    \end{equation}
\end{conjecture}
\end{mdframed}
\vspace{3mm}
In flat space, if the gauge coupling depends on the moduli, the extremality factor will be modified. In our work, for concreteness, we restrict the scope of the discussion to an $\AdS_D$ Einstein--Maxwell theory, without referencing the quantum gravity uplift from which it could have been derived. We expect the conjectured bound to receive modifications in the presence of additional fields, which we leave for future investigation. 

Our Conjecture \ref{conj:WGCinAdS} is the first formulation of the WGC in AdS space by a concrete exploration of the possibility for an extremal RN-AdS black hole to decay --- a key motivation when the WGC was originally formulated in flat space \cite{Arkani-Hamed:2006emk}. Our conjecture agrees, as it should, with the formulation of the WGC in flat space, leading to the familiar WGC when taking the limit $\eta \rightarrow 0$,
\begin{equation}\label{eq:WGCinflatspace-intro}
    \frac{g_\UoD{D}\mf{q}}{\mf{m}}\ge \sqrt{\gamma}\frac{1}{M_\PlD{D}^{\frac{D-2}{2}}}\coma
\end{equation}
The modification to the WGC bound that we obtain depends on the ratio of the horizon radius of the RN-AdS black hole and the AdS length. The bound is monotonic with respect to $\eta$ and is maximized for the largest possible RN-AdS black holes with $r_h \gg \ell_\AdSD{D}$. In this limit, Conjecture \ref{conj:WGCinAdS} requires \eqref{eq:chargetomassforconjecture-maximum}, whose inequality is stronger than \eqref{eq:WGCinflatspace-intro} by a factor of $\sqrt{\gamma}$. As a sharp reader might have noticed, the above bound seems to strictly exclude BPS states, and thus a supersymmetric setup. We will address this puzzle in \cref{sec:commentsonBPSbound,sec:conclusions}.

\subsubsection*{Structure of the Paper}
The paper is structured as follows. In Section \ref{sec:review}, we review the WGC in flat space and its formulation with multiple $\U(1)$ charges, i.e., the convex hull condition (CHC) \cite{Cheung:2014vva}. We discuss the connection between the WGC and the RFC in Section \ref{sec:WGCvsRFC}, obtaining the usual formulation of the WGC as a repulsive force condition on the particle at the horizon of a RN black hole in Section \ref{sec:NoForceInMinko}. Following the review, we discuss the WGC in AdS space. Starting from a brief recap of existing AdS proposals in Section \ref{sec:WGCinAdS-previous}, we then introduce the geometry of an extremal RN-AdS black hole in Section \ref{sec:RN-AdSBH}. 

We derive our AdS WGC bound in Section \ref{sec:WGCwithSchwinger} by extending the results in \cite{Lin:2024jug} to $D$-dimensional RN-AdS black hole spacetime. The bound is determined as the condition on the charge-to-mass ratio of the charged particle for the Schwinger effect to take place. In Section \ref{sec:WGCfromBFBound}, we demonstrate the connection of the Schwinger effect to the Breitenlohner--Freedman (BF) instability condition \cite{Breitenlohner:1982bm,Gubser:2008px,Gubser:2008pf,Denef:2009tp} for the charged particle in the vicinity of the extremal black hole horizon, which exhibits a local geometry of $\AdS_2\times S^{D-1}$. Next, in \cref{sec:RFCinAdS,sec:NFC-BHvspart}, we confirm \eqref{eq:chargetomassforconjecture} in light of the repulsive force condition in AdS, by requiring the charged particle to feel a repulsive force at the horizon of an extremal RN-AdS black hole. Agreement between the bounds based on the three arguments is shown, and we conclude with our proposal of the AdS WGC Conjecture.

The generalization of Conjecture \ref{conj:WGCinAdS} in the case of RN-AdS black holes and particles charged under multiple U(1) gauge fields is discussed in Section \ref{sec:CHCinAdS}, leading to the convex hull version as given by Conjecture \ref{conj:CHCAdS}. In Section \ref{sec:commentsonBPSbound}, we comment on the application of our proposal of WGC in AdS to theories with extended supersymmetry. We conclude in Section \ref{sec:conclusions} and discuss possible uplifts and potential tests of our conjecture in string theory.

\section{Weak Gravity Conjecture in Minkowski}
\label{sec:review}

In this section, we review the original formulation of the WGC \cite{Arkani-Hamed:2006emk}, as well as its extension in the presence of multiple abelian gauge symmetries. For a more detailed overview of the literature, we also refer to the original works on tower WGC \cite{Heidenreich:2016aqi,Montero:2016tif,Andriolo:2018lvp} and the reviews \cite{Brennan:2017rbf,Palti:2019pca,Palti:2020mwc,vanBeest:2021lhn,Grana:2021zvf,Harlow:2022ich,Agmon:2022thq,Reece:2023czb,Rudelius:2024mhq,Montero:2024qml}.

\subsubsection*{Weak Gravity Conjecture for Charged Particles}

Let us consider the following Einstein--Maxwell action in Einstein frame:
\begin{equation}\label{eq:Schargedparticles}
    S \supset \frac{M_{\PlD{D}}^{D-2}}{2}\int_{\mathcal{M}_D}  R\star\ID -\frac{1}{2g_{\UoD{D}}^2}\int_{\mathcal{M}_D} F_2\wedge \star F_2\coma
\end{equation}
where $M_\PlD{D}$ is the $D$-dimensional Planck mass, and $g_\UoD{D}$ is the $\U(1)$ gauge coupling.\footnote{In our convention, $F_{p+2}\wedge \star F_{p+2}=\frac{1}{(p+2)!}F_{\mu_1\ldots \mu_{p+2}}F^{\mu_1\ldots \mu_{p+2}}$, moreover recall that in a $D$-dimensional theory, the gauge coupling has mass dimensions $\frac{4-D}{2}$. If such an action is obtained from some compactification of a higher-dimensional theory, then it will be a function of the moduli defining the volumes of the compactification space; however, for the purpose of our paper, we assume that it does not depend on the moduli.} 

We can consider a particle of mass $\mf{m}$ charged under the $\U(1)$ gauge symmetry with charge
\begin{equation}
    \mf{q} = \frac{1}{g_\UoD{D}^2}\int_{\mathcal{M}_{D-2}}\star F_2\fstop
\end{equation}
The WGC \cite{Arkani-Hamed:2006emk} postulates that for any $\U(1)$ gauge field theory coupled to gravity, there must exist an object whose charge-to-mass ratio is larger than the charge-to-mass ratio of an extremal black hole charged under the same gauge fields, i.e.
\begin{equation}\label{eq:WGCgeneric}
    \frac{\mf{q}^2}{\mf{m}^2}\geq \left.\frac{\mathcal{Q}^2}{\mathcal{M}^2}\right|_\ttiny{ext.}\fstop
\end{equation}
One can write the charge-to-mass ratio of an extremal black hole in terms of the Planck mass of the $D$-dimensional theory, obtaining
\begin{equation}\label{eq:extremalgamma}
    \left.\frac{g_\UoD{D}^2\mathcal{Q}^2}{\mathcal{M}^2}\right|_\ttiny{ext.} \equiv \gamma \frac{1}{M_\PlD{D}^{D-2}}\coma
\end{equation}
where $\gamma$ is the extremality factor. If the gauge coupling does not depend on the moduli, the extremality factor is the same as the charge-to-mass ratio of a $D$-dimensional Reissner--Nordstr\"om black hole \cite{Heidenreich:2015nta,Harlow:2022ich}, i.e.
\begin{equation}
\label{eq:gammafactorDef}
    \gamma = \frac{D-3}{D-2}\fstop
\end{equation}
Using \eqref{eq:extremalgamma}, the WGC becomes
\begin{equation}
\label{eq:DdimWGC}
    \frac{g_\UoD{D}^2\mf{q}^2}{\mf{m}^2}\geq \gamma \frac{1}{M_\PlD{D}^{D-2}}\fstop
\end{equation}

The WGC is often introduced together with its magnetic version, which sets the scale at which the EFT is expected to be valid in the weak-coupling limit of the gauge theory. This scale is set by the mass of the magnetic monopoles in Maxwell theory, which is proportional to the inverse of the gauge coupling. The magnetic WGC then imposes that 
\begin{equation}
    \Lambda_\WGCD{D}^2\lesssim g_{\UoD{D}}^2 M_\PlD{D}^{D-2}\fstop
\end{equation}

\subsubsection*{The Convex Hull Weak Gravity Conjecture}

In the presence of multiple $\U(1)$ gauge theories, there exists a generalization of the WGC that goes by the name of Convex Hull Condition (CHC) \cite{Cheung:2014vva}. For every massive particle in the spectrum with mass $\mf{m}_i$ charged under multiple $\U(1)_i$, with charge $\mf{q}_i$, we introduce the vector of charge-to-mass ratios as
\begin{equation}\label{eq:zvector}
    \vec{\mf{z}}_i = \frac{M_\PlD{D}^{\frac{D-2}{2}}}{m_i}\gamma^{-1/2}\left(g_\UoD{D, 1}\mf{q}_1,\ldots,g_\UoD{D, n}\mf{q}_n\right)\fstop
\end{equation}
The CHC implies that the convex hull formed by the $\vec{\mf{z}}_i$ vectors of all the multiparticle states must include the unit ball.
Via dimensional reduction or duality, one obtains the CHC for axions \cite{Rudelius:2015xta,Brown:2015iha,Brown:2015lia}
which has been used to constrain multi-axion inflation models.

\subsection{Weak Gravity Conjecture as Repulsive Force Conjecture}
\label{sec:WGCvsRFC}

The WGC is motivated by the requirement that gravity is weaker than other forces. However, the formulation above does not involve a comparison between the gravitational force and the others. Rather, it imposes a condition on the particle spectrum such that they are required to be superextremal. If the only forces involved in the theory are the gravitational and electromagnetic ones, it is possible to connect the two ideas. As has been studied in \cite{Heidenreich:2019zkl}, superextremal particles will feel
the repulsion of the electromagnetic interaction as stronger than the gravitational attraction. This led to the formulation of the Repulsive Force Conjecture (RFC), which states that in any effective field theory coupled to gravity with a $\U(1)$ gauge field, there exists a self-repulsive charged particle \cite{Heidenreich:2019zkl}. 

The equivalence between the WGC and the RFC is properly true only in the absence of massless scalar fields. However, the two conjectures still remain related even in the presence of moduli, since the objects with vanishing long-range self-force are also those satisfying some version of the WGC (see e.g. \cite{Heidenreich:2019zkl,Lee:2018spm,Lee:2018urn,Lee:2019wij,Klaewer:2020lfg,Gendler:2020dfp,Cota:2022yjw,Cota:2022maf,FierroCota:2023bsp,Heidenreich:2024dmr}).

The purpose of this paper is to use the relationship between the RFC and the WGC in the absence of a massless scalar to extend the WGC in an Einstein--Maxwell theory in AdS background. We then find the condition under which there exist a range of separation between the black hole and the particle where the net force on the particle is non-attractive, and it will provide the lower bound for a repulsive force formulation in AdS background. Assuming that RFC and WGC are related in the absence of massless scalars for any background, the results will provide the version of the WGC for backgrounds that are not asymptotically flat. 

In the following, we will first recover the relationship between the WGC and the RFC in flat space, and we will repeat the computation with a non-zero cosmological constant in Section \ref{sec:WGCinAdS}.

\subsection{Repulsive Force Condition in Minkowski}
\label{sec:NoForceInMinko}

In \cite{Heidenreich:2019zkl}, the repulsive (attractive) force is seen when calculating the energy between the interacting particles. We will instead analyze the force through geodesics, computing the force between holes and particles, and showing that the repulsive force condition so derived is consistent with the dynamics of fields that allow for the decay of extremal black holes.

In theories with diffeomorphism invariance, the notion of force and acceleration can be subtle. This can be overcome by working in static coordinates so that a zero force or acceleration indicates a fixed proper separation. The force will be determined from the coordinate acceleration of a probe particle in spacetime containing either another particle or a black hole.

We start with the Einstein--Maxwell action in $D=d+1$ dimensions
\beq \label{eq:EMinMink}
I = \frac{M_{\PlD{D}}^{D-2}}{2}\int_{\mathcal{M}_D}  R\star\ID -\frac{1}{2g_{\UoD{D}}^2}\int_{\mathcal{M}_D} F_2\wedge \star F_2\,,
\eeq
where $F_2=dA_1$ is the 2-form field strength associated with the U(1) gauge symmetry and $g_\UoD{D}$ is the coupling strength. The quantized charge enclosed in a codimension-2 surface is defined as
\beq
\mc{Q}=\frac{1}{g_\UoD{D}^2}\int_{\mc{M}_{D-2}} \star \,F_2\fstop
\eeq
We introduce the gauge field corresponding to a charged black hole or particle as
\beq \label{gauge_field}
A_1=\left(-\frac{g_{\UoD{D}}^2\mc{Q}}{(D-3)\omega_{D-2}r^{D-3}}+C\right)dt\coma
\eeq
where $C$ is a constant that we will fix later when we zoom in to the near-horizon region \cite{Hartnoll:2008kx}. The physics is not affected by the constant since it depends only on the field strength, i.e., the derivative of the gauge field. The metric can be computed from the Einstein equations and it is given by
\beq
ds^2=-f_\MinkD{D}(r)dt^2+\frac{dr^2}{f_\MinkD{D}}+r^2d\Omega_{D-2}^2\coma
\eeq
where
\beq
f_\MinkD{D}(r)=1-\frac{2M}{r^{D-3}}+\frac{Q^2}{r^{2(D-3)}}\fstop
\label{eq:fMinkD}
\eeq
While it is convenient to work with the scaled parameters $M$ and $Q$, they are related to the ADM mass and quantized charge by \cite{Hawking:1995fd,Myers:1986un}
\begin{equation}
    M = \frac{\gamma}{(D-3)\omega_{D-2}}\frac{\mc{M}}{M_\PlD{D}^{D-2}}\coma Q = \frac{\sqrt{\gamma}}{(D-3)\omega_{D-2}}\frac{g_\UoD{D}\mc{Q}}{M_\PlD{D}^{\frac{D-2}{2}}}\coma
    \label{eq:mqinMQ}
\end{equation}
where $\gamma$ is the extremality factor introduced in \eqref{eq:gammafactorDef}.\footnote{Note that $Q$ is properly defined only for $D>3$ and we will assume $D>3$ for all the computations in this work.}

This metric describes a black hole when there exists an event horizon defined by $f_\MinkD{D}(r_h)=0$ and when the black hole size is larger than the Planck length $\ell_\PlD{D}\equiv M_\PlD{D}^{-1}$. When the mass of the black hole is sufficiently lower than the Planck mass, the metric can be used to describe the spacetime containing a point particle at a large distance compared to the Schwarzschild radius.

Generally, \eqref{eq:fMinkD} admits two horizons $r_\pm$, but we are interested in the extremal case where $r_+=r_-=r_h$. At this horizon, 
\begin{equation}\label{eq:QMinhorizonMinko}
Q = M = r_h^{D-3}\coma
\end{equation} 
that in terms of \eqref{eq:mqinMQ} reduces to \eqref{eq:extremalgamma}. Given this condition, we fix $C$ in \eqref{gauge_field} as
\begin{equation} \label{eq:chemical_potential}
    C = \frac{g_{\UoD{D}}^2\mc{Q}}{(D-3)\omega_{D-2}r_h^{D-3}}\coma
\end{equation}
so that the gauge field is zero at the horizon. This gauge choice admits a smooth and finite gauge potential when moving close to the black hole horizon.
Now let us consider the motion of a charged particle with mass $\mf{m}$ and electric charge $\mf{q}$ in the background of an extremal RN black hole. The particle will have an action
\beq
S_{\mf{m},\mf{q}}=\int d\tau \left[ \mf{m}\sqrt{g_{\mu\nu}(\xi)\dot{\xi}^\mu\dot{\xi}^\nu}+\mf{q}A_\mu(\xi)\dot{\xi^\mu}\right]\coma
\eeq
where $\xi^\mu$ is the worldline of the point particle and $\dot{\xi}^\mu\equiv \frac{d\xi^\mu}{d\tau}$. The equations of motion of the particle in the presence of an RN black hole read
\beq
\frac{d^2\xi^\mu}{d\tau^2}+\Gamma^\mu_{\rho\sigma}\frac{d\xi^\rho}{d\tau}\frac{d\xi^\sigma}{d\tau}=\frac{\mf{q}}{\mf{m}}g^{\mu\rho}F_{\rho\nu}\frac{d\xi^\nu}{d\tau}\fstop 
\label{eq:geodesicEq}
\eeq
The idea is to find the condition under which the particle is repelled by the black hole at the horizon, so that it is not reabsorbed by the black hole.\footnote{While this seems to be a classical argument for the decay of extremal black holes, one can check that the effective potential is monotonically growing if the repulsive force condition is not met, shutting off even the possibility of quantum tunneling.} In order to do so, we consider the force on the particle when it is at rest, i.e. $\frac{dx^\nu}{d\tau}=\left(f_\MinkD{D}^{-\frac{1}{2}}, 0, \cdots, 0\right)$. 
This leads to the radial equation of motion, defining the force density
\beq \label{eq:FrMink}\bal
F(r) \equiv \ddot{r}&=-\Gamma^1_{00}f_\MinkD{D}^{-1}+\frac{\mf{q}}{\mf{m}}f_\MinkD{D}^{\frac{1}{2}}A_t'\\
&=-\frac{1}{2}f'_\MinkD{D}+\frac{\mf{q}}{\mf{m}}\sqrt{f_\MinkD{D}}A'_t\fstop
\eal \eeq
We therefore require that $F(r)\ge 0$ at $r\in[r_h, r_h+\delta]$ for some positive $\delta$. Notice that $f_\MinkD{D}(r_h)=f'_\MinkD{D}(r_h)=0$, since the extremal horizon is defined as the Killing horizon with zero surface gravity, so, by \eqref{eq:FrMink}, we also have $F(r_h)=0$. We only need to require that the derivative of the force is positive, namely $F'(r_h)\ge 0$, i.e.
\begin{equation}\label{eq:FpcondMinko}
  \left.  F'(r)\right|_{r\rightarrow r_h} = \left.\left(-\frac{1}{2}f''_\MinkD{D}+\frac{\mf{q}}{\mf{m}}\frac{f'_\MinkD{D}}{2\sqrt{f_\MinkD{D}}}A'_t+\frac{\mf{q}}{\mf{m}}\sqrt{f_\MinkD{D}}A''_t\right)\right|_{r\rightarrow r_h}\geq 0\fstop
\end{equation}
Notice that the expression above must be interpreted as a limit for $r\rightarrow r_h$ because the ratio $\frac{f'_\MinkD{D}}{\sqrt{f_\MinkD{D}}}$ is apparently infinite. However, we can expand $f_\MinkD{D}(r)$ and its derivatives in a neighborhood of $r_h$ to obtain 
\beq \label{eq:fMinkoexpa}\bc
f_\MinkD{D}(r)=(D-3)^2\frac{(r-r_h)^2}{r_h^2}+\mathcal{O}((r-r_h)^3)\\
f'_\MinkD{D}(r)=2(D-3)^2\frac{r-r_h}{r_h^2}+\mathcal{O}((r-r_h)^2)\\
f''_\MinkD{D}(r)=2(D-3)^2\frac{1}{r_h^2}+\mathcal{O}(r-r_h)
\ec \eeq
noticing that away from the horizon, the ratio $\frac{f'_\MinkD{D}}{\sqrt{f_\MinkD{D}}}$ is finite. This means that we can satisfy \eqref{eq:FpcondMinko} by requiring
\beq\label{eq:chargeconditioninMinko}
\frac{\mf{q}}{\mf{m}}\ge\left.\frac{\sqrt{f_\MinkD{D}}f''_\MinkD{D}}{A'_t f'_\MinkD{D}}\right|_{r\rightarrow r_h}\fstop
\eeq
The derivative of the gauge potential in \eqref{gauge_field} at the horizon is
    \begin{equation}\label{eq:derivativegauge_field}
  A_t'(r_h) = \frac{g_{\UoD{D}}^2\mc{Q}}{\omega_{D-2}r_h^{D-2}}\coma
\end{equation}
so that, plugging \cref{eq:fMinkoexpa,eq:derivativegauge_field} into \eqref{eq:chargeconditioninMinko}, we obtain
\begin{equation}
    \frac{\mf{q}}{\mf{m}}\geq \frac{(D-3)\omega_{D-2}}{g_\UoD{D}^2\mathcal{Q}}r_h^{D-3} = \frac{\sqrt{\gamma}}{g_\UoD{D}M_\PlD{D}^{\frac{D-2}{2}}}\coma
\end{equation}
where in the last step we have expressed $r_h$ in terms of the charge $Q$ in \eqref{eq:mqinMQ}. We have then recovered the WGC bound in \eqref{eq:DdimWGC} as
\beq
\frac{g_\UoD{D}\mf{q}}{\mf{m}}\ge \sqrt{\gamma}\frac{1}{M_\PlD{D}^{\frac{D-2}{2}}}\equiv \left.\frac{g_\UoD{D}\mathcal{Q}}{\mathcal{M}}\right|_\text{\tiny ext.}\coma \label{eq:WGCfromBHinMink}
\eeq
as expected. More importantly, this result is also obtained from the RFC in the absence of scalars. For convenience later, we express this bound in terms of the charge-to-mass ratio $\mf{z}$, defined as in \eqref{eq:zvector}, i.e.,
\begin{equation}\label{eq:chargetomassratioparticle}
    \mf{z} \equiv \frac{g_\UoD{D}\mf{q}M_\PlD{D}^{\frac{D-2}{2}}}{\sqrt{\gamma}\mf{m}}\coma
\end{equation}
so that \eqref{eq:WGCfromBHinMink} is equivalent to 
\begin{equation}
\mf{z}^2 = \frac{\gamma^{-1} g^2_\UoD{D}\mf{q}^2 M_\PlD{D}^{D-2}}{\mf{m}^2}\geq 1\fstop
\end{equation}

\section{Weak Gravity Conjecture in Anti-de Sitter}
\label{sec:WGCinAdS}

In flat space, the WGC bound can be seen from the black hole extremality condition, black hole decay condition, and particle self-repulsiveness, which all lead to the same bound \cite{Heidenreich:2019zkl,Heidenreich:2020upe}. Particle bounds obtained from the different arguments undergo non-trivial modifications in the presence of a cosmological constant and should be carefully revisited. In this section, we argue that extremal black hole decay due to dynamics of charged fields remains the core of formulating the WGC in AdS.\footnote{It was observed in \cite{Horowitz:2022mly} that perturbations can be singular at the horizon of an extremal AdS black hole. This is in line with the possibility explored in our work. Extremal RN-AdS black holes should decay, for instance through Schwinger effect, so that the black hole relaxes to a non-extremal one free of divergence.} It is consistent with the appropriate generalization of the RFC between extremal black holes and particles in AdS space. In fact, in AdS space, a repulsive force gives rise to the possibility of a particle created by the black hole not being reabsorbed, thus leading to its decay. The agreement between the conditions due to the quantum effect of particle creation by extremal black holes and the classical picture of the particle being repelled by the black hole indicates the appropriateness of using them as the basis of our generalization of the WGC in AdS space.

In the following, we first review the existing speculations on the WGC in AdS and derive the extremality condition for RN-AdS black holes. We will then motivate our proposal for the AdS WGC by analyzing the Schwinger effect of RN-AdS black holes, the near-horizon BF instability condition, and the requirement for a particle to be repelled by the black hole.

\subsection{Previous Proposals for WGC in AdS}
\label{sec:WGCinAdS-previous}

In \cite{Nakayama:2015hga}, it was proposed that in $\AdS_D$ with length scale $\ell_\AdSD{D}$, a more natural quantity to compare the charge of a particle with is not the mass of the particle, but its energy at rest, i.e. $\Delta \,\ell_\AdSD{D}^{-1}$. The relation between $\mf{m}$ and $\Delta$ depends on the dimension of the space and the spin of the particle, however, for scalar fields, it is 
\begin{equation}\label{eq:conformaldimension}
    \Delta = \frac{D-1}{2}+\sqrt{\frac{(D-1)^2}{4}+\ell_\AdSD{D}^2\mf{m}^2}\fstop
\end{equation}
One of the requirements to formulate the WGC considered in \cite{Nakayama:2015hga} was that it reduces to its usual formulation in the flat space limit, i.e. \eqref{eq:DdimWGC} for $\ell_\AdSD{D}\rightarrow \infty$. This leads to the proposal that in $\AdS_D$ there must exist a particle such that
\begin{equation}
\label{eq:WGCinAdSproposedbefore}
    g_\UoD{D}^2 \mf{q}^2 \geq \gamma \frac{1}{M_\PlD{D}^{D-2}} \Delta^2 \ell_\AdSD{D}^{-2}\fstop
\end{equation}
While $\Delta$ seems a more natural quantiy from the perspective of the CFT dual, in the context of AdS/CFT correspondence, the formulation of the WGC in AdS as in \eqref{eq:WGCinAdSproposedbefore} is justified, mainly, by the requirement that it reduces to the usual formulation of the WGC in the flat space limit. This requirement may be considered necessary but not sufficient to make it more justified than other expressions that have the same flat space limit. In the following, we first consider the regime where $\ell_\AdSD{D}^2\mf{m}^2 \gg 1$, so that \eqref{eq:WGCinAdSproposedbefore} reduces to the usual WGC expressed in terms of $\mf{m}$. However, in Section \ref{sec:commentsonBPSbound}, we return to this proposal of WGC in the context of EFTs with extended supersymmetry. 

\subsection{Reissner--Nordstr\"om AdS Black Holes} 
\label{sec:RN-AdSBH}

To study the WGC in AdS space, we include the cosmological constant term in \eqref{eq:EMinMink}. The action then reads
\beq
I = \frac{M_\PlD{D}^{D-2}}{2}\int \left[R  -\frac{(D-1)(D-2)}{\ell^2_\AdSD{D}}\right]\star \ID- \frac{1}{2g_\UoD{D}^2}\int F_2 \wedge \star F_2\,,
\eeq
where $\ell_\AdSD{D}$ is the AdS length scale. The cosmological constant is identified as $\Lambda_\AdSD{D}=-\frac{(D-1)(D-2)}{\ell^2_\AdSD{D}}$.

In AdS space, the gauge field remains \eqref{gauge_field} and the RN-AdS metric follows from the Einstein equations
\beq
\label{RN-AdS_metric}
ds^2=-f_\AdSD{D}(r)dt^2+\frac{dr^2}{f_\AdSD{D}}+r^2d\Omega^2_{D-2}\coma
\eeq
with
\beq
f_\AdSD{D}(r) = 1-\frac{2M}{r^{D-3}}+\frac{Q^2}{r^{2(D-3)}}+\frac{r^2}{\ell^2_\AdSD{D}}\,.
\label{eq:fAdSD}
\eeq
The parameters are related to the physical quantities as in \eqref{eq:mqinMQ}. 

Similarly to the RN metric in flat space, $f_\AdSD{D}(r)$ admits two roots, $r_\pm$. When the two roots are real, i.e.,
\begin{equation}
    \frac{D-1}{D-3}r_+^{2(D-2)}+\ell^2_\AdSD{D}r_+^{2(D-3)}\geq Q^2\ell^2_\AdSD{D}\fstop
    \label{eq:horizoncondition}
\end{equation}
the spacetime is free of naked singularity\footnote{It is interesting to notice that if the black hole admits a supersymmetric embedding, a BPS black hole, i.e. satisfying $M = Q$, can never be extremal (see e.g. \cite{Romans:1991nq,Caldarelli:1998hg,Chamblin:1999tk,Hristov:2011ye,Hristov:2012bd}). However, for supersymmetric black holes, the metric has 
\begin{equation}
    f_\ttiny{SUSY}(r) = \left(1-\frac{Q}{r^{D-3}}\right)^2+\frac{r^2}{\ell^2_\AdSD{D}}\coma
\end{equation}
which is strictly positive everywhere, leading to a naked singularity for $r = 0$. This means that purely electric extremal RN-AdS black holes cannot be BPS.\label{foot:SUSYAdSBH}} and the larger positive root $r_+$ of $f_\AdSD{D}(r)$ represents the location of the black hole event horizon. In addition, if \eqref{eq:horizoncondition} is saturated, the black hole is extremal. The valid parameter range of the RN-AdS black hole can therefore be written as $M\geq M_\ttiny{ext.}(Q,\ell_\AdSD{D})$. By saturating \eqref{eq:horizoncondition}, we can express the mass and the charge of a black hole in terms of the horizon radius $r_h$, leading to
\beq\label{eq:MQwrhinAdS} \bc
M=r_h^{D-3}\left(1+\frac{D-2}{D-3}\eta^2\right)\coma\\
Q^2=r_h^{2(D-3)}\left(1+\frac{D-1}{D-3}\eta^2\right)\coma
\ec \eeq
where we have introduced 
\begin{equation}\label{eq:eta}
    \eta = \frac{r_h}{\ell_\AdSD{D}}\fstop
\end{equation} 
This is the main difference compared to the Minkowski case in \eqref{eq:QMinhorizonMinko}. In the Minkowski case, the extremality bound was imposing that charge and mass of the black hole are equal, i.e. $M=Q$, resembling some sort of BPS condition in supersymmetric set-ups. However, this is no longer true in a curved background, where extremality is simply the condition in which the inner and outer horizons of the RN-AdS black hole agree. This has interesting consequences when we try to apply \eqref{eq:DdimWGC} and compute the charge-to-mass ratio for an extremal RN-AdS black hole, obtaining
\begin{equation}
    \left.\frac{g_\UoD{D}^2\mc{Q}^2}{\mc{M}^2}\right|_\ttiny{ext.}  = \gamma\frac{1}{M_\PlD{D}^{D-2}}\frac{1+\frac{D-1}{D-3}\eta^2}{\left(1+\frac{D-2}{D-3}\eta^2\right)^2} = \gamma \frac{1}{M_\PlD{D}^{D-2}}\left(1-\eta^2+\mathcal{O}\left(\eta^4\right)\right)\fstop
    \label{eq:extremalBHinAdSratio}
\end{equation}
Substituting this constraint into \eqref{eq:DdimWGC}, it is tempting to use the AdS BH extremal bound as the AdS version of WGC. This AdS bound as defined would be weaker than the WGC bound in flat space, since the RHS of \eqref{eq:extremalBHinAdSratio} is generally smaller than 1, and approaches zero for large black holes. Moreover, it reduces to the WGC in flat space when we take the limit $\eta\rightarrow \infty$. However, we will show in Section \ref{sec:WGCwithSchwinger}, that if one bounds the particle spectrum by the AdS BH extremality, it is not guaranteed that an extremal RN-AdS black hole can decay. A stronger bound, based on the black hole decay argument, is needed. We will derive this new bound in the following sections.

\subsection{The Weak Gravity Conjecture from Schwinger Effect}
\label{sec:WGCwithSchwinger}

The original motivation for the WGC was to ensure the decay of extremal black holes. In the case of cold black holes, i.e. $T_H\ll \mf{m}$, with $\mf{m}$ being the mass of the emitted particle, the specific channel responsible for the decay process is by Schwinger effect \cite{Schwinger:1951nm}. A full analysis of the spatial profile and threshold behavior of the production rate was conducted in \cite{Lin:2024jug}, confirming that in flat space, the dynamics of matter fields around the black hole are consistent with the Cosmic Censorship Conjecture (CCC) and WGC,\footnote{See also \cite{Crisford:2017gsb} where it was shown numerically in an axial-symmetric set-up that particles satisfying the WGC can save the CCC through condensation.} namely only particles with charge-to-mass ratio higher than the charge-to-mass ratio of the corresponding extremal black hole can lead to charged emission without leaving behind a naked singularity. The result is also in agreement with the study of the Schwinger effect in $\AdS_2$ space \cite{Pioline:2005pf}, which can be understood as a local analysis in the $\AdS_2\times S^2$ horizon geometry.
While it happens that in flat space black hole extremality and black hole decay indicate the same bound on the charged particle spectrum, this no longer holds in the presence of a negative cosmological constant. 

In this section, we are going to review some relevant results obtained in \cite{Lin:2024jug}. We will generalize the computation to $D$-dimensional AdS space and focus on the critical charge-to-mass ratio of the particle, below which the Schwinger effect is switched off.

We recall that the Schwinger rate is related to the vacuum-vacuum amplitude in the presence of the background gauge field. The amplitude is calculated by integrating out the charged field, resulting in an effective action. The effective action is then evaluated using the worldline instanton method, where the instanton action provides the exponential suppression factor and the second-order fluctuation provides the prefactor. The term worldline reflects the nature of the instantons being charged particle worldlines in the curved Euclidean spacetime. Further, the worldlines have to satisfy the periodic boundary condition, i.e., starting and ending at the same point.

The threshold value for the charge-to-mass ratio lies in the prefactor of the effective action and is inversely proportional to the square root of the determinant of the path fluctuations and the traverse speed along the Euclidean path. The general analysis in the full black hole spacetime requires numerical computations, but the threshold value can be found analytically near the black hole horizon where the electric field is strongest. The proper distance of the worldline, fixing the total proper time, grows when the charge-to-mass ratio of the particle is lowered towards the critical value, effectively sending the prefactor of the Schwinger production rate to zero. When the critical value is reached, the worldline instanton degenerates and ceases to exist below the critical charge-to-mass ratio. Therefore, we can identify the threshold value as the point where periodic worldline instantons cannot be found.

Having established the connection between the threshold charge-to-mass ratio value and the vanishing of periodic worldline instantons, we move on to obtain the precise threshold value in $D$-dimensional AdS space. The worldline instanton solutions satisfy a set of equations similar to the geodesic equation of a charged particle coupled to a Maxwell field in curved space, except for having a Euclidean signature. In \cite{Lin:2024jug}, it was found that the radial equation is
\beq
\dot{r}=\pm a\sqrt{f_\AdSD{D}(r)-\frac{\mf{q}^2}{\mf{m}^2}\left[A_t(r)-\omega\right]^2}\coma
\eeq
where $a$ is a normalization factor that sets the scale of the derivative of $r$ and $\omega$ determines the radial range of the worldline. The worldline must be periodic, meaning that it must have turn-around points where $\dot{r} = 0$. This requires that
\begin{equation}
    h(r) = f_\AdSD{D}(r)-\frac{\mf{q}^2}{\mf{m}^2}\left[A_t(r)-\omega\right]^2\coma
\end{equation}
has a zero at some $r_*>r_h$ outside the horizon for some $\omega$ and that $h(r)\ge0$ for $r\in [r_*, r_*+\delta]$ for some $\delta>0$. In this case, the worldline instanton will be a loop in the Euclidean space where $h(r)$ is positive and with $r=r_*$ as one of the turn-around points. The bound on the charge-to-mass ratio is determined by the instantons near the horizon region of the black hole, motivating the change of coordinates\footnote{It is important that the constant term in the gauge field is properly chosen so that the gauge field vanishes at the horizon for this limit to be smooth and finite when taking $\epsilon\rightarrow 0$, which requires the choice of the constant in \eqref{gauge_field} to be \eqref{eq:chemical_potential}.} $r=r_h+\epsilon \rho$ and $\omega\rightarrow \epsilon \omega$, so that $h(\rho)$ is cast into a quadratic form
\beq
h(\rho)=\frac{f''_\AdSD{D}(r_h)}{2}\epsilon^2\rho^2-\frac{\mf{q}^2}{\mf{m}^2} \epsilon^2{A'_t}^2(r_h)(\rho-\omega)^2\fstop
\eeq
The requirement on $h(r)$ passes to the requirement on $h(\rho)$ having roots at $\rho>0$ for some choice of $\omega$. Observe that $h(\omega)>0$ and $h(0)\le 0$, therefore we only need to set the coefficient of the quadratic term to be negative for any positive $\omega$ to ensure two non-negative roots of $h(\rho)$, i.e., 
\beq
\frac{\mf{q}^2}{\mf{m}^2}\geq\left.\frac{f''_\AdSD{D}}{2{A'_t}^2}\right|_{r\rightarrow r_h}\fstop
\label{eq:boundfromSchwinger}
\eeq
In order to compute the RHS of \eqref{eq:boundfromSchwinger}, we proceed in the same way as in Section \ref{sec:NoForceInMinko}. We notice that at the horizon, $f_\AdSD{D}(r_h)=f'_\AdSD{D}(r_h) =0$, but $f''_\AdSD{D}(r_h)$ is constant. We can expand around the horizon $f_\AdSD{D}(r)$ even though, for the moment, we are only interested in its second derivative:
\beq \label{f_expansion}\bc
f_\AdSD{D}(r)=\left[(D-3)^2+(D-1)(D-2)\eta^2\right]\frac{(r-r_h)^2}{r_h^2}+\mathcal{O}((r-r_h)^3)\\
f'_\AdSD{D}(r)=2\left[(D-3)^2+(D-1)(D-2)\eta^2\right]\frac{r-r_h}{r_h^2}+\mathcal{O}((r-r_h)^2)\\
f''_\AdSD{D}(r)=2\left[(D-3)^2+(D-1)(D-2)\eta^2\right]\frac{1}{r_h^2}+\mathcal{O}(r-r_h)
\ec \eeq
where we have introduced again $\eta$ as in \eqref{eq:eta}. On the other hand, the derivative of the gauge field is still \eqref{eq:derivativegauge_field} because the AdS background has not affected the definition of the gauge field. 
The RHS of \eqref{eq:boundfromSchwinger} is thus evaluated to be
\beq \bal
\left.\frac{f''_\AdSD{D}}{2{A'_t}^2}\right|_{r\rightarrow r_h}&=\frac{\omega_{D-2}^2\left[(D-3)^2+(D-1)(D-2)\eta^2\right]}{g_\UoD{D}^2\mc{Q}^2}r_h^{2(D-3)}\fstop
\eal \eeq
On the other hand, we know from \cref{eq:mqinMQ,eq:MQwrhinAdS} that
\begin{equation}\label{eq:rhintermsofQ}
    r_h^{D-3} = \frac{Q}{\sqrt{1+\frac{D-1}{D-3}\eta^2}} = \frac{1}{\sqrt{1+\frac{D-1}{D-3}\eta^2}}\frac{\sqrt{\gamma}}{(D-3)\omega_{D-2}}\frac{g_\UoD{D}\mc{Q}}{M_\PlD{D}^{\frac{D-2}{2}}}\fstop
\end{equation}
Plugging back the result into \eqref{eq:boundfromSchwinger} we obtain
\beq \label{eq:Schwinger_bound-almost}
\frac{\mf{q}^2}{\mf{m}^2}\geq \frac{\gamma}{g_\UoD{D}^2M_\PlD{D}^{D-2}}\left(\frac{1+\frac{(D-1)(D-2)}{(D-3)^2}\eta^2}{1+\frac{D-1}{D-3}\eta^2}\right)\fstop
\eeq
We can, once again, introduce the charge-to-mass ratio $\mf{z}$ as in \eqref{eq:chargetomassratioparticle}, obtaining the bound
\begin{equation}\label{eq:Schwinger_bound}
    \mf{z}^2 \geq \frac{1+\frac{(D-1)(D-2)}{(D-3)^2}\eta^2}{1+\frac{D-1}{D-3}\eta^2}\equiv \mathcal{R}(\eta)^2\fstop
\end{equation}
We notice that the bound is drastically different from \eqref{eq:extremalBHinAdSratio}. It reduces to 
\begin{equation}
    \mf{z}^2 \geq 1\coma
\end{equation}
in the flat space limit, but generally $1\le \mathcal{R}(\eta) < \sqrt{\frac{D-2}{D-3}}$. The bound is maximized when the RN-AdS black hole is much larger than the AdS length, i.e. $r_h \gg \ell_\AdSD{D}$ or equivalently $\eta \rightarrow \infty$, where
\begin{equation}
    \mf{z}^2 \geq \frac{D-2}{D-3} =\gamma^{-1}\coma
\end{equation}
strictly larger than $1$ for $D>3$. This is the first time we explicitly encounter the bound that leads to our Conjecture \ref{conj:WGCinAdS}. This is the condition on the emitted particle that would allow the decay of an extremal RN-AdS black hole, so that, in principle, could be sufficient to conclude the necessity of Conjecture \ref{conj:WGCinAdS}, based on the original motivation of the WGC. However, this will not be the only time we obtain this bound, and, in fact, we are going to derive it multiple times from different perspectives.

\subsection{Relation to BF Bound Instability}
\label{sec:WGCfromBFBound}

The Schwinger effect of an extremal black hole can also be understood as a consequence of instability in the near-horizon $\AdS_2$ geometry of the black hole, i.e., the charged particle has an effective mass below the BF bound. This is the same instability that triggers the scalar condensation around AdS black holes, sometimes referred to as the charged superradiance instability \cite{Denef:2009tp, Chamblin:1999tk,Gubser:2008pf, Hartnoll:2008kx,Basu:2010uz}.

The BF bound dictates the stability of a neutral scalar field in AdS space. The focus of this paper is the decay of RN-AdS black holes and not the $D$-dimensional AdS space, so the particles we consider will not violate the BF bound in $\AdS_D$. However, the extremal black hole has a near-horizon geometry of $\AdS_2\times S^{D-2}$ and a scalar field becomes unstable at the horizon when the 2-dimensional BF bound is violated
\beq\label{eq:BFinAdS2}
m_\text{\tiny eff.}^2\le -\frac{1}{4\ell_{\ttiny{AdS}_2}^2}\coma
\eeq
where for a charged scalar field, the effective mass is shifted from the bare mass in the near-horizon region \cite{Gubser:2008px} to
\beq \label{eq:eff_mass}
m^2_\text{\tiny eff.}=\left.\left(\mf{m}^2-\mf{q}^2f^{-1}_\AdSD{D}A_t^2\right)\right|_{r\rightarrow r_h}\fstop
\eeq
Let us notice again that the expression above must be interpreted as a limit for $r\rightarrow r_h$, since both  $A_t$ and $f_\AdSD{D}(r)$ are zero for $r=r_h$. However, we have seen in \eqref{f_expansion} that around the horizon for an extremal RN-AdS, $f_\AdSD{D}(r)$ can be approximated with
\begin{equation}
    \left.f_\AdSD{D}(r)\right|_{r\rightarrow r_h} = \frac{1}{L^2}(r- r_h)^2\coma
\end{equation}
where we can identify $L^{-2}=\frac{f''_\AdSD{D}(r_h)}{2}$. On the other hand, by expanding \eqref{gauge_field} around $r_h$, we can rewrite it as
\begin{equation}
    A_t=-\frac{g_{\UoD{D}}^2\mc{Q}}{\omega_{D-2}r_h^{D-2}} (r-r_h)+\mathcal{O}((r-r_h)^2)\coma
\end{equation}
where we have used the fact that the constant $C$ is fixed to be \eqref{eq:chemical_potential}, so that the gauge potential is zero at the horizon.\footnote{One should think of this as a gauge fixing condition that is finite in the near-horizon coordinates. Physics does not depend on the gauge since the dynamics is concerned with derivatives of the potential, nor does thermodynamics because the chemical potential of the horizon is given by $\mu\equiv A_{t}(r_h)-A_{t}(\infty)$. A different choice of gauge will lead to a divergent effective mass at the black hole horizon, which requires additional care when analyzing the instability.} This means that in the ratio in \eqref{eq:eff_mass}, when expanding in power series of $r$ around $r_h$ there will be a single non-zero contribution that remains constant, while all the other contributions go to zero for $r\rightarrow r_h$. The effective mass as of \eqref{eq:eff_mass} is then
\beq\label{eq:instabilitycondition}
m^2_\text{\tiny eff.}=\mf{m}^2-\frac{g^4_\UoD{D}\mf{q}^2 \mc{Q}^2 L^2}{\omega_{D-2}^2r_h^{2(D-2)}}\fstop
\eeq
Now, we need to understand the meaning of the length $L$ in terms of the $\AdS_2$ geometry of the black hole. In order to do so, we make a change of coordinates
\beq \bc
r=r_h+\frac{\epsilon}{y}\\
t=\frac{L^2}{\epsilon}d\tau\coma
\ec \eeq
so that the near-horizon limit corresponds to $\epsilon\rightarrow 0$. With these coordinates, the RN-AdS metric transforms to
\beq
ds^2=L^2\left(\frac{-d\tau^2+dy^2}{y^2}\right)+r_h^2 d\Omega_{D-2}^2\coma
\eeq
which has the geometry of $\AdS_2\times S^{D-2}$. We find that the $\AdS_2$ length scale is precisely $L$,
\beq\label{eq:AdS_length2d}
\ell_{\ttiny{AdS}_2}=L=\frac{f''_\AdSD{D}}{2}=\frac{r_h}{\sqrt{(D-3)^2+(D-1)(D-2)\eta^2}}\fstop
\eeq

Now we return to the instability condition in \eqref{eq:instabilitycondition}. When the state with mass $\mf{m}$ 
is well described by a point particle, i.e., when the particle's Compton wavelength is much smaller than the black hole size $\mf{m}^{-1}\ll r_h\sim \ell_\AdSD{2}$, the RHS of \eqref{eq:instabilitycondition} is subleading compared to the mass shift due to the gauge field. Plugging \eqref{eq:AdS_length2d} into \eqref{eq:instabilitycondition}, and replacing $r_h$ with \eqref{eq:rhintermsofQ}, the effective mass becomes
\beq \bal
m_\text{\tiny eff.}^2& = \mf{m}^2-\frac{g^2_\UoD{D}\mf{q}^2M_\PlD{D}^{D-2}}{\gamma} \frac{1+\frac{D-1}{D-3}\eta^2}{1+\frac{(D-1)(D-2)}{(D-3)^2}\eta^2}\fstop
\eal \eeq
This mass must satisfy \eqref{eq:BFinAdS2}, i.e.,
\beq
\mf{m}^2-\frac{g^2_\UoD{D}\mf{q}^2M_\PlD{D}^{D-2}}{\gamma} \frac{1+\frac{D-1}{D-3}\eta^2}{1+\frac{(D-1)(D-2)}{(D-3)^2}\eta^2}\le -\frac{1}{4\ell_\AdSD{2}^2}\coma
\eeq
or analogously
\begin{mdframed}[backgroundcolor=white, shadow=true, shadowsize=4pt,shadowcolor=seccolor,
roundcorner=6pt]
\begin{remark}\label{remark:BFbound}
\begin{equation}
\label{eq:BF_with_O1}
\frac{g^2_\UoD{D}\mf{q}^2M_\PlD{D}^{D-2}}{\gamma \mf{m}^2} \geq \frac{1+\frac{(D-1)(D-2)}{(D-3)^2}\eta^2}{1+\frac{D-1}{D-3}\eta^2}\left(1+\frac{1}{4\mf{m}^2\ell_\AdSD{2}^2}\right)\fstop
\end{equation} 
\end{remark}
\end{mdframed}
\vspace{3mm}

The result in \eqref{eq:BF_with_O1} is the bound proposed in Conjecture \ref{conj:WGCinAdS} that also applies to states with masses $\mf{m}\ell_\AdSD{2}\ll 1$. Since, for the moment, we are considering $\mf{m}\ell_\AdSD{2}\gg 1$,\footnote{We will consider situations when the second term cannot be neglected in Section \ref{sec:commentsonBPSbound}, where we will discuss states whose Compton wavelength is comparable with the AdS length.} the above bound reduces to
\beq\label{eq:zfromdecay}
\frac{g^2_\UoD{D}\mf{q}^2M_\PlD{D}^{D-2}}{\gamma \mf{m}^2}\equiv \mf{z}^2  \ge \frac{1+\frac{(D-1)(D-2)}{(D-3)^2}\eta^2}{1+\frac{D-1}{D-3}\eta^2}+\mathcal{O}\left(\frac{1}{\mf{m}^2r_h^2}\right)\fstop
\eeq
The first term gives exactly the RHS of \eqref{eq:Schwinger_bound}. The additional correction term is understood as the correction from the wavelike nature of the charge particle. It is only relevant when the Compton wavelength of the particle is comparable to the black hole size, i.e., when $\mf{m}r_h\sim 1$. 

The BF instability condition agrees with the analysis of the Schwinger effect of the extremal RN-AdS black hole as expected, and it represents the second time Conjecture \ref{conj:WGCinAdS} appears. In the next section, we will consider a third and last scenario, in which we will find the conditions for which a charged particle in a RN-AdS black hole background feels a repulsive force at the horizon. This last computation will provide the last compelling evidence for Conjecture \ref{conj:WGCinAdS}.

\subsection{Repulsive Force Condition in AdS Space}
\label{sec:RFCinAdS}

In flat space, two extremal charged objects interacting only through gravity and a U(1) gauge field experience cancellation of the two forces. In \cite{Palti:2017elp}, the RFC is proposed as an alternative formulation of the WGC. Subsequent papers examined the forces, including scalar-mediated force, between identical charged particles at long range \cite{Heidenreich:2020upe}. These studies showed a connection between repulsive force and the black hole extremality condition in flat space. As we have shown, extremality in AdS space differs from the black hole decay requirement already. A natural question that arises is whether the repulsive force condition agrees with either of the two bounds.

In fact, before addressing this question, one has to specify what it means to say that particles are repulsive. First, there is no longer a good notion of ``long range" in AdS space where there exists a confining potential due to the negative cosmological constant. The force between two massive particles will always be attractive at a sufficiently large separation. The repulsive force condition can be argued by considering the binding energy between charged states, as in \cite{Andriolo:2022hax}. However, the crucial point is to understand what kind of charged objects to consider when discussing the WGC bound. We adhere to the idea that WGC allows for extremal black holes to decay; therefore, we will consider the interaction between an extremal RN-AdS black hole and a charged particle. We hereby formulate the notion of repulsiveness between the black hole and the particle as the following:
\begin{mdframed}[backgroundcolor=white, shadow=true, shadowsize=4pt,shadowcolor=seccolor,
roundcorner=6pt]
A charged black hole and a particle are said to be repulsive if there exists a range where the force between them is non-attractive. 
\end{mdframed}
\vspace{3mm}
The force is determined by evaluating the geodesic motion of the charged particle. When the radial acceleration of the particle is non-negative, the force is non-attractive. In the following, we will compute the force for the case of an extremal RN-AdS black hole.

\subsection{Repulsive Force between an RN-AdS Black Hole and a Particle}
\label{sec:NFC-BHvspart}

The computation of the force of an extremal black hole on a particle has been previously described in Section \ref{sec:NoForceInMinko}. Here we will follow the steps we described in Section \ref{sec:NoForceInMinko} and generalize them to AdS space. We should stress that, for an extremal black hole, imposing the repulsive force condition on the black hole horizon is equivalent to our formulation in the previous section.

Formally, the radial equation of motion for the particle is still those obtained from \eqref{eq:geodesicEq}, but with $f_\MinkD{D}(r)$ in \eqref{eq:fMinkD} replaced by $f_\AdSD{D}(r)$ in \eqref{eq:fAdSD}. This gives the force density 
\begin{equation} \label{eq:AdS_force}
    F(r) =-\frac{1}{2}f'_\AdSD{D}+\frac{\mf{q}}{\mf{m}}f^{\frac{1}{2}}_\AdSD{D}A'_t\coma
\end{equation}
with the gauge field defined as in \eqref{gauge_field}. The repulsive force condition requires that $F'(r_h)\geq 0$, namely
\begin{equation}
\left.  F'(r)\right|_{r\rightarrow r_h} = \left.\left(-\frac{1}{2}f''_\AdSD{D}+\frac{\mf{q}}{\mf{m}}\frac{f'_\AdSD{D}}{2\sqrt{f_\AdSD{D}}}A'_t+\frac{\mf{q}}{\mf{m}}\sqrt{f_\AdSD{D}}A''_t\right)\right|_{r\rightarrow r_h}\geq 0\coma    
\end{equation}
which is, once again, satisfied at the horizon when
\begin{equation}
    \frac{\mf{q}}{\mf{m}}\ge\left.\frac{\sqrt{f_\AdSD{D}}f''_\AdSD{D}}{A'_t f'_\AdSD{D}}\right|_{r\rightarrow r_h}\fstop
    \label{eq:CompWGCboundfromNoforceinAdS}
\end{equation}
However, here is where the difference between the flat space and the AdS space becomes important. The extremal solution does not correspond anymore to the case where $M=Q$, but instead, the mass and charge of the RN-AdS black hole are related to the horizon radius by \eqref{eq:MQwrhinAdS}. This is also reflected when we expand \eqref{eq:fAdSD} around the horizon radius $r_h$, as in \eqref{f_expansion}. We can compute the RHS of \eqref{eq:CompWGCboundfromNoforceinAdS} as we explained in Section \ref{sec:NoForceInMinko} and similarly to the computation in Section \ref{sec:WGCfromBFBound}. In fact, nothing conceptually changes and the RHS of \eqref{eq:CompWGCboundfromNoforceinAdS} is 
\begin{equation}
    \left.\frac{\sqrt{f_\AdSD{D}}f''_\AdSD{D}}{A'_t f'_\AdSD{D}}\right|_{r\rightarrow r_h} = \frac{\omega_{D-2}(D-3)\sqrt{1+\frac{(D-1)(D-2)}{(D-3)^2}\eta^2}}{g_{\UoD{D}}^2\mc{Q}}r_h^{D-3}\fstop
\end{equation}
By expressing $r_h$ in terms of the charges as in \eqref{eq:rhintermsofQ}, we obtain 
\begin{equation}
    \frac{g_\UoD{D}\mf{q}}{\mf{m}}\ge \sqrt{\gamma}\frac{1}{M_\PlD{D}^{\frac{D-2}{2}}} \sqrt{\frac{1+\frac{(D-1)(D-2)}{(D-3)^2}\eta^2}{1+\frac{D-1}{D-3}\eta^2}}\coma
\end{equation}
or analogously 
\begin{equation}\label{eq:WGCfromBHinAdS}
    \mf{z} \geq \sqrt{\frac{1+\frac{(D-1)(D-2)}{(D-3)^2}\eta^2}{1+\frac{D-1}{D-3}\eta^2}}\equiv \mathcal{R}(\eta)\fstop
\end{equation}
We immediately notice that this bound matches with the one obtained in \cref{sec:WGCwithSchwinger,sec:WGCfromBFBound}, and when saturated it corresponds to the minimal charge-to-mass ratio that a particle must have in order for the extremal RN-AdS black hole to decay. It also reduces to \eqref{eq:DdimWGC} in the flat space limit $\eta\rightarrow 0$ since
\beq
\mathcal{R}(\eta)=1+\frac{D-1}{2(D-3)^2}\eta^2 + O(\eta^3)\fstop
\eeq
However, as we anticipated, this is not the bound we would have obtained if we generalized the WGC using \eqref{eq:extremalBHinAdSratio}. This computation provides further evidence that this is the correct generalization of the WGC in curved background and to the formulation of Conjecture \ref{conj:WGCinAdS}. The function $\mathcal{R}(\eta)$ in the RHS of \eqref{eq:WGCfromBHinAdS} is a monotonic function of $\eta$, ranging between $1\leq \mathcal{R}(\eta)< \gamma^{-1/2}$, which then leads to a bound that depends on the size of the RN-AdS black hole. However, the strongest bound is obtained by considering the largest possible RN-AdS black hole, with $r_h\gg \ell_\AdSD{D}$ (or analogously $\eta\rightarrow\infty$). If we require then a stronger version of Conjecture \ref{conj:WGCinAdS}, in which the charge-to-mass ratio of a particle must allow the largest RN-AdS black hole to decay, then
\beq
\mf{z}\ge \gamma^{-1/2}\fstop
\eeq

\subsection{A Convex Hull Weak Gravity Conjecture in AdS}
\label{sec:CHCinAdS}

Having established the WGC in AdS as in Conjecture \ref{conj:WGCinAdS}, the first natural extension is to consider the decay condition for a RN-AdS black hole charged under multiple $\U(1)$s. The way in which we are going to extend the conjecture is by requiring that a probe particle is repelled by the RN-AdS black hole at the horizon, as we did in the previous section for a single $\U(1)$. The action, in the case of $n$ $\U(1)$, is given by\footnote{\label{foot:unmixedU1}One can consider a more general Lagrangian with kinetic mixing among the $\U(1)$s, but the upshot of our computation does not get modified. The following action can be regarded as the action obtained by diagonalizing the kinetic matrix, although, in that case, the charges are generically not quantized. However, the quantization of the charges does not play any role in our computation.} 
\beq
I = \frac{M_\PlD{D}^{D-2}}{2}\int \left[R  -\frac{(D-1)(D-2)}{\ell^2_\AdSD{D}}\right]\star \ID- \frac{1}{2}\sum_{i=1}^n\frac{1}{g_\UoD{D, i}^2}\int F_2^{(i)} \wedge \star F_2^{(i)}\,,
\eeq
and we choose the gauge field to be 
\beq
A_t=-\frac{1}{(D-3)\omega_{D-2}}\sum_{i=1}^n \frac{g_{\UoD{D, i}}^2\mc{Q}_i}{r^{D-3}}+C\coma
\eeq
which is a direct generalization of \eqref{gauge_field}. The solution to the RN-AdS metric remains given by \eqref{RN-AdS_metric}, but this time \eqref{eq:fAdSD} is
\begin{equation}
    f_\AdSD{D}(r) = 1-\frac{2M}{r^{D-3}}+\frac{Q_T^2}{r^{2(D-3)}}+\frac{r^2}{\ell^2_\AdSD{D}}\,,
\end{equation}
where we have introduced
\begin{equation}
    Q_T^2 = \sum_i Q_i^2\coma
\end{equation}
where $Q_i$ are related to the quantized charge $\mathcal{Q}_i$ by \eqref{eq:mqinMQ} for each $\U(1)$. When we consider an extremal RN-AdS black hole, the values of the parameters $M$ and $Q_T$ are still related to the horizon radius by \eqref{eq:MQwrhinAdS}, where we replace $Q$ by $Q_T$. We can consider now the motion of a particle of mass $\mf{m}$ charged under the $n$ $\U(1)$s with charges $\mf{q}_i$ in this background and find the condition for which such a particle is repelled by the black hole at the horizon. The force density is the generalization of \eqref{eq:AdS_force}, but is still given by 
\beq
F(r) =-\frac{1}{2}f'_\AdSD{D}(r)+f^{\frac{1}{2}}_\AdSD{D}(r)\sum_{i=1}^n \frac{\mf{q_i}}{\mf{m}}\frac{d}{dr}A^{(i)}_t\fstop
\eeq
Once again we require that at the horizon the force is repulsive, i.e. $F'(r_h)\geq 0$, which means, since $f_\AdSD{D}(r)|_{r\rightarrow r_h}\rightarrow 0$, that we must require
\begin{equation}
    \left.-\frac{1}{2}f''_\AdSD{D}(r)+\frac{f'_\AdSD{D}(r)}{2\sqrt{f_\AdSD{D}(r)}}\sum_{i=1}^n \frac{\mf{q_i}}{\mf{m}}\frac{d}{dr}A^{(i)}_t\right|_{r\rightarrow r_h} \geq 0\coma
\end{equation}
or equivalently, since $f''_\AdSD{D}(r_h) \neq 0$, 
\begin{equation}
    \left.\frac{f'_\AdSD{D}(r)}{f''_\AdSD{D}(r)\sqrt{f_\AdSD{D}(r)}}\sum_{i=1}^n \frac{\mf{q_i}}{\mf{m}}\frac{d}{dr}A^{(i)}_t\right|_{r\rightarrow r_h} \geq 1\fstop
\end{equation}
The computation, then, proceeds as in the previous section, with the right modification, eventually, one obtains\footnote{From \eqref{eq:partialCHCcondition}, we can read how to generalize the constraint if instead we start from a more general Lagrangian containing kinetic mixing, i.e.
\begin{equation}
    I \supset -\frac{1}{2}\sum_{i,j=1}^n\mathcal{G}_{ij}\int F_2^{(i)}\wedge\star F_2^{(j)}\coma
\end{equation}
where $\mathcal{G}_{ij}$ are the components of a gauge kinetic matrix $\mathcal{G}$. \eqref{eq:partialCHCcondition} generalizes as
\begin{equation}\label{eq:generalCHCcondition}
    \sqrt{\frac{1+\frac{D-1}{D-3}\eta^2}{1+\frac{(D-1)(D-2)}{(D-3)^2}\eta^2}}\frac{M_\PlD{D}^{\frac{D-2}{2}}\gamma^{-1/2}}{\mf{m}}\frac{\sum_{i,j=1}^n \mf{q}_i \mathcal{G}^{ij}\mathcal{Q}_i}{\sqrt{\sum_{a,b=1}^n \mc{Q}_a \mathcal{G}^{ab}\mc{Q}_b}}\geq 1\coma
\end{equation}
where $\mathcal{G}^{ab}$ are the components of $\mathcal{G}^{-1}$. Everything that follows can be analogously discussed also using \eqref{eq:generalCHCcondition}, however, to keep the discussion simpler, we will consider only unmixed $\U(1)$s as we explained in Footnote \ref{foot:unmixedU1}.}
\begin{equation}\label{eq:partialCHCcondition}
    \sqrt{\frac{1+\frac{D-1}{D-3}\eta^2}{1+\frac{(D-1)(D-2)}{(D-3)^2}\eta^2}}\frac{M_\PlD{D}^{\frac{D-2}{2}}\gamma^{-1/2}}{\mf{m}}\frac{\sum_{i=1}^n g_\UoD{D, i}^2 \mf{q}_i \mathcal{Q}_i}{\sqrt{\sum_{j=1}^n g_\UoD{D, j}^2\mc{Q}_j^2}}\geq 1\fstop
\end{equation}
We can introduce the charge-to-mass ratio vector for the particle as in \eqref{eq:chargetomassratioparticle} so that the bound becomes
\begin{equation}
    \frac{\sum_{i=1}^n g_\UoD{D, i} \mf{z}_j \mathcal{Q}_j}{\sqrt{\sum_{j=1}^n g_\UoD{D, j}^2\mc{Q}_j^2}} \geq  \sqrt{\frac{1+\frac{(D-1)(D-2)}{(D-3)^2}\eta^2}{1+\frac{D-1}{D-3}\eta^2}}\fstop
\end{equation}
Finally, if we introduce the charge vector for the black hole 
\begin{equation}
    \vec{\mathcal{Q}} = \left(g_\UoD{D, 1}\mc{Q}_1,\ldots,g_\UoD{D, n}\mc{Q}_n\right)\coma 
\end{equation}
then we obtain
\begin{equation}\label{eq:CHCWGCinAdS}
    \frac{\vec{\mf{z}} \cdot \vec{\mathcal{Q}}}{|\vec{\mathcal{Q}}|} \geq  \sqrt{\frac{1+\frac{(D-1)(D-2)}{(D-3)^2}\eta^2}{1+\frac{D-1}{D-3}\eta^2}}\fstop
\end{equation}

In order to understand \eqref{eq:CHCWGCinAdS}, let us consider the flat space limit, i.e. $\eta\rightarrow 0$. In this case, \eqref{eq:CHCWGCinAdS} reduces to 
\begin{equation}
     \frac{\vec{\mf{z}} \cdot \vec{\mathcal{Q}}}{|\vec{\mathcal{Q}}|} = \vec{\mf{z}} \cdot \vec{\mathcal{Z}} \geq 1\coma
     \label{eq:RFCinFlatspacemulticharge}
\end{equation}
where we have introduced the charge-to-mass ratio of an extremal RN black hole being
\begin{equation}\label{eq:ZforBH}
    \vec{\mathcal{Z}} = \frac{\gamma^{-1/2}\vec{\mc{Q}}M_\PlD{D}^{\frac{D-2}{2}}}{\mc{M}}\coma
\end{equation}
that has unit norm in flat space. The norm of $\vec{\mathcal{Z}}$ is the same for any choice of charges $\vec{\mathcal{Q}}$ and mass $\mathcal{M}$ provided that 
\begin{equation}\label{eq:extremalityinFS}
    \mathcal{M}^2 = \gamma^{-1}M_\PlD{D}^{D-2} |\vec{\mathcal{Q}}|^2\coma
\end{equation}
which is the extremality condition in flat space. 

We provide an equivalent interpretation of the convex hull condition in \cite{Cheung:2014vva} using the picture of black hole decay. We first emphasize that different extremal black holes specified by the vector $\vec{\mc{Z}}$ should all be allowed to decay through the set of charged particles in the theory. If there is only a single charged particle, it cannot satisfy \eqref{eq:RFCinFlatspacemulticharge} for all $\vec{\mc{Z}}$, and therefore it will not be sufficient to decay all black holes in the theory. The resolution is to ensure that there is a set of charged particles $\{\vec{\mf{z}}_i\}$ such that for any $\vec{\mc{Z}}$, \eqref{eq:RFCinFlatspacemulticharge} is satisfied by at least one particle. Under this condition, all extremal black holes can emit the particles satisfying \eqref{eq:RFCinFlatspacemulticharge} and become non-extremal, where Hawking radiation can lead to its further decay.

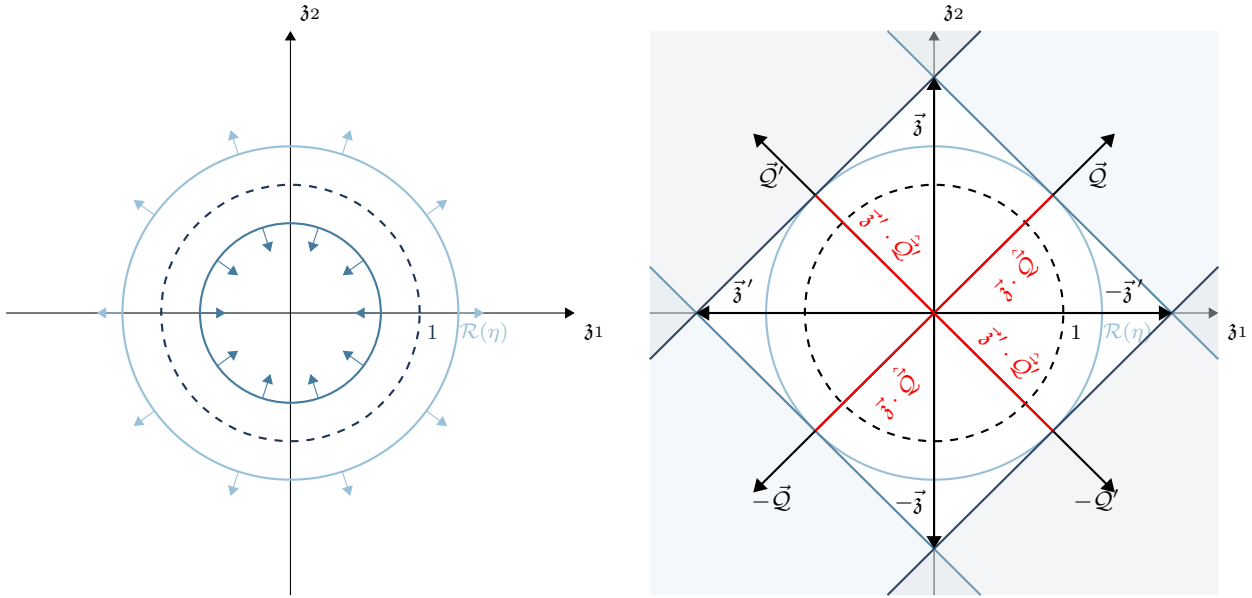
\begin{figure}[!htp]
    \centering
    \begin{subfigure}[t]{0.49\textwidth}
    \centering
    \begin{tikzpicture}[baseline=0,font=\footnotesize, scale=1.7]
        \draw[-Triangle] (-2.2,0) -- node[below right, pos=1] {$\mf{z}_1$}(2.2,0);
        \draw[-Triangle] (0,-2.2) -- node[above right, pos=1] {$\mf{z}_2$}(0,2.2);
        \draw[thick,dashed,prcolor] (0,0) circle (1);
        \draw[thick,seccolor] (0,0) circle  (0.7);
        \draw[thick,tercolor] (0,0) circle (1.3);
        \node[prcolor] at (1.1,-0.15) {\scriptsize$1$};
        \node[tercolor] at (1.5,-0.15) {\scriptsize$\mathcal{R}(\eta)$};
        \foreach \x in {1,...,10}{\draw[-Triangle, seccolor] ({\x*360/10}:0.7) -- ({\x*360/10}:0.5);}
        \foreach \x in {1,...,10}{\draw[-Triangle, tercolor] ({\x*360/10}:1.3) -- ++({\x*360/10}:0.2);}
    \end{tikzpicture}
        \caption{Plot of {\color{seccolor}{$\mathcal{R}_\ttiny{ext.}(\eta)$}} and {\color{tercolor}{$\mathcal{R}(\eta)$}} as a function of $\eta$. 
        The unit circle is obtained for $\mathcal{R}_\ttiny{ext.}(0)=\mathcal{R}(0)$, while, for $\eta \rightarrow \infty$, ${\color{seccolor}{\mathcal{R}_\ttiny{ext.}(\eta)}}\rightarrow 0$ and  ${\color{tercolor}{\mathcal{R}(\eta)}}\rightarrow \gamma^{-1/2}$.
        }
    \label{fig:extremalityineta}
    \end{subfigure}\hfill
    \begin{subfigure}[t]{0.49\textwidth}
    \begin{tikzpicture}[baseline=0,font=\footnotesize, scale=1.7]
        \draw[-Triangle] (-2.2,0) -- node[below right, pos=1] {$\mf{z}_1$}(2.2,0);
        \draw[-Triangle] (0,-2.2) -- node[above right, pos=1] {$\mf{z}_2$}(0,2.2);
        \draw[thick,dashed] (0,0) circle (1);
        \draw[thick,tercolor] (0,0) circle (1.3);
        \node at (1.1,-0.15) {\scriptsize$1$};
        \node[tercolor] at (1.5,-0.15) {\scriptsize$\mathcal{R}(\eta)$};
        
        \draw[thick,seccolor] (-0.36,2.2) -- (2.2,-0.36);
        \fill[seccolor!20,nearly transparent] (-0.36,2.2) -- (2.2,-0.36) -- (2.2,2.2) -- (-0.36,2.2);
        \draw[thick,-Triangle] (0,0) -- node[pos=0.8,left] {$\vec{\mf{z}}$} (0,1.84);
        \draw[thick,-Triangle] (0,0) -- node[pos=0.9,below] {$\vec{\mathcal{Q}}$} (1.4,1.4);
        \draw[thick,prhigh] (0,0) -- node[pos=0.5,below, sloped] {$\vec{\mf{z}}\cdot\hat{\vec{\mathcal{Q}}}$} (0.92,0.92);

        \draw[thick,seccolor] (0.36,-2.2) -- (-2.2,0.36);
        \fill[seccolor!20,nearly transparent] (0.36,-2.2) -- (-2.2,0.36) -- (-2.2,-2.2) -- (0.36,-2.2);
        \draw[thick,-Triangle] (0,0) -- node[pos=0.8,left] {$-\vec{\mf{z}}$} (0,-1.84);
        \draw[thick,-Triangle] (0,0) -- node[pos=0.9,below] {$-\vec{\mathcal{Q}}$} (-1.4,-1.4);
        \draw[thick,prhigh] (0,0) -- node[pos=0.5,below, sloped] {$\vec{\mf{z}}\cdot\hat{\vec{\mathcal{Q}}}$} (-0.92,-0.92);

        \draw[thick,prcolor] (0.36,2.2) -- (-2.2,-0.36);
        \fill[prcolor!20,nearly transparent] (0.36,2.2) -- (-2.2,-0.36) -- (-2.2,2.2) -- (0.36,2.2);
        \draw[thick,-Triangle] (0,0) -- node[pos=0.8,above] {$\vec{\mf{z}}\,'$} (-1.84,0);
        \draw[thick,-Triangle] (0,0) -- node[pos=0.9,below] {$\vec{\mathcal{Q}}'$} (-1.4,1.4);
        \draw[thick,prhigh] (0,0) -- node[pos=0.5,above, sloped] {$\vec{\mf{z}}\,'\cdot\hat{\vec{\mathcal{Q}}}'$} (-0.92,0.92);

        \draw[thick,prcolor] (-0.36,-2.2) -- (2.2,0.36);
        \fill[prcolor!20,nearly transparent] (-0.36,-2.2) -- (2.2,0.36) -- (2.2,-2.2) -- (-0.36,-2.2);
        \draw[thick,-Triangle] (0,0) -- node[pos=0.8,above] {$-\vec{\mf{z}}\,'$} (1.84,0);
        \draw[thick,-Triangle] (0,0) -- node[pos=0.9,below] {$-\vec{\mathcal{Q}}'$} (1.4,-1.4);
        \draw[thick,prhigh] (0,0) -- node[pos=0.5,above, sloped] {$\vec{\mf{z}}\,'\cdot\hat{\vec{\mathcal{Q}}}'$} (0.92,-0.92);
    \end{tikzpicture}
        \caption{Note that the charge vector $\vec{\mathcal{Q}}$ increases with $\eta\rightarrow \infty$, but \eqref{eq:CHCWGCinAdS} requires that the projection of $\vec{\mf{z}}$ on $\vec{\mathcal{Q}}$ is at least equal to $\mathcal{R}(\eta)$. $\vec{\mathcal{Q}}$ and $\vec{\mathcal{Q}}'$ are two possible charges of RN-AdS such that $|\vec{\mathcal{Q}}|=|\vec{\mathcal{Q}}'|$. In order to satisfy \eqref{eq:CHCWGCinAdS} for any choice of $\vec{\mathcal{Q}}$, at least two states $\vec{\mf{z}}$ and $\vec{\mf{z}}\,'$ (and their anti-particles) are necessary.}
    \label{fig:CHCAdS-example}
    \end{subfigure}
    \caption{Schematic representation of the CHC in AdS. In Figure \ref{fig:extremalityineta}, we show how the charge-to-mass ratio of an extremal RN-AdS black hole and the decay bound we propose in \eqref{eq:CHCWGCinAdS} differs from  unity when $r_h\gg \ell_\AdSD{D}$. In Figure \ref{fig:CHCAdS-example}, we give an example of how to satisfy the CHC in AdS provided the existence of at least two particles (and anti-particles) satisfying \eqref{eq:CHCWGCinAdS}.}
    \label{fig:CHCAdS}
\end{figure}

If we now consider a non-trivial cosmological constant, the bound becomes \eqref{eq:CHCWGCinAdS}, where we can still express it in terms of $\vec{\mathcal{Z}}$ whenever the horizon of the extremal, now, RN-AdS black hole is finite, i.e.
\begin{equation}\label{eq:CHCWGCinAdSwithZBH}
    \frac{\vec{\mf{z}} \cdot \vec{\mathcal{Z}}}{|\vec{\mathcal{Z}}|} \geq  \sqrt{\frac{1+\frac{(D-1)(D-2)}{(D-3)^2}\eta^2}{1+\frac{D-1}{D-3}\eta^2}}\fstop
\end{equation}
The main difference is in the expression of $\vec{\mathcal{Z}}$ compared to the RN black hole in flat space. While its definition is still given by \eqref{eq:ZforBH}, the extremality condition for a RN-AdS is not given by \eqref{eq:extremalityinFS} but rather by \eqref{eq:extremalBHinAdSratio}. This means that increasing the horizon of the black hole, will reduce the norm of $\vec{\mathcal{Z}}$, while the RHS of \eqref{eq:CHCWGCinAdS} increases. In the limit of $\eta \rightarrow \infty$, 
\begin{equation}
    |\vec{\mathcal{Z}}| =  \left.\frac{\gamma^{-1/2}g_\UoD{D}\mc{Q}M_\PlD{D}^{\frac{D-2}{2}}}{\mc{M}}\right|_\ttiny{ext.} \equiv \mathcal{R}_\ttiny{ext.}(\eta) = \sqrt{\frac{1+\frac{D-1}{D-3}\eta^2}{\left(1+\frac{D-2}{D-3}\eta^2\right)^2}} \stackrel{\eta\rightarrow \infty}{\longrightarrow} 0\coma
\end{equation}
while 
\begin{equation}
    \mathcal{R}(\eta) = \sqrt{\frac{1+\frac{(D-1)(D-2)}{(D-3)^2}\eta^2}{1+\frac{D-1}{D-3}\eta^2}}\stackrel{\eta\rightarrow \infty}{\longrightarrow} \sqrt{\frac{D-2}{D-3}}=\gamma^{-1/2}\fstop
\end{equation}
We show, schematically, this effect in Figure \ref{fig:extremalityineta}. This is the essential difference between a RN-AdS black hole and a RN black hole in flat space. However, the logic that leads to the CHC in flat space, can still be pursued by using \eqref{eq:CHCWGCinAdS}. By introducing 
\begin{equation}
    \hat{\vec{\mathcal{Q}}} = \frac{\vec{\mathcal{Q}}}{|\vec{\mathcal{Q}}|}\coma
\end{equation}
\eqref{eq:CHCWGCinAdS} means that for any choice of charges $\vec{\mathcal{Q}}$, there must exist a particle whose charge-to-mass ratio lies at or above the tangent plane to a ball of radius $\mathcal{R}(\eta)$. The scalar product in \eqref{eq:CHCWGCinAdS} is well-defined even in the limit of $\eta \rightarrow \infty$, solving the ambiguity with the CHC in AdS due to the shrinking of the ball with radius $\mathcal{R}_\ttiny{ext.}(\eta)$. The conclusion is that in AdS, there must exist as many particles with charge-to-mass ratios $\vec{\mf{z}}_i$ such that their convex hull contains the ball of radius $\mathcal{R}(\eta)$. The result is illustrated in Figure \ref{fig:CHCAdS-example}. Given this interpretation, we are led to propose the following conjecture: 

\begin{mdframed}[backgroundcolor=white, shadow=true, shadowsize=4pt,shadowcolor=seccolor,
roundcorner=6pt]
    \begin{conjecture}[AdS Convex Hull Weak Gravity Conjecture]\label{conj:CHCAdS}
    Given $n$ $\U(1)$ gauge fields coupled to Einstein--Maxwell gravity, then, for every direction of the charge vector 
    \begin{equation}
        \vec{\mathcal{Q}} = \left(g_\UoD{D, 1}\mc{Q}_1,\ldots,g_\UoD{D, n}\mc{Q}_n\right)\coma 
    \end{equation}
    of a multi-charged RN-AdS black hole, there must exist at least one massive particle  satisfying 
    \begin{equation}
       \frac{\vec{\mf{z}_i} \cdot \vec{\mathcal{Q}}}{|\vec{\mathcal{Q}}|} \geq \sqrt{\frac{1+\frac{(D-1)(D-2)}{(D-3)^2}\eta^2}{1+\frac{D-1}{D-3}\eta^2}}\coma \text{with } \eta = \frac{r_h}{\ell_\AdSD{D}}\coma
    \end{equation}
    where $r_h$ is the horizon of an extremal RN-AdS black hole and $\ell_\AdSD{D}$ is the $\AdS_D$ scale length and each $\vec{\mf{z}_i}$ is defined as
    \begin{equation}
         \vec{\mf{z}_i} = \frac{M_\PlD{D}^{\frac{D-2}{2}}}{m_i}\gamma^{-1/2}\left(g_\UoD{D, 1}\mf{q}_{1,i},\ldots,g_\UoD{D, n}\mf{q}_{n,i}\right)\fstop
    \end{equation} 
    The value $\gamma$ is the extremality factor for a RN black hole in flat space, i.e. 
    $\gamma  = \frac{D-3}{D-2}$.
    
    Analogously, the convex hull formed by $\{\vec{\mf{z}}_i\}$ must contain a ball of radius 
    \begin{equation}
        \mathcal{R}(\eta)= \sqrt{\frac{1+\frac{(D-1)(D-2)}{(D-3)^2}\eta^2}{1+\frac{D-1}{D-3}\eta^2}}\fstop
    \end{equation}
    Considering the largest possible RN-AdS black hole, i.e. $r_h\gg \ell_\AdSD{D}$, the radius of the ball becomes 
    \begin{equation}
       \mathcal{R}_\infty= \gamma^{-1/2}\fstop
    \end{equation}
\end{conjecture}
\end{mdframed}
\vspace{3mm}

Our argument for the AdS Convex Hull is different from that given in \cite{Cheung:2014vva}, where the black hole is assumed to decay completely into a set of particles and one considers the convex hull generated by all the decay products. Here, we only require the extremal black hole to decay into a non-extremal one; therefore, the particles responsible for the convex hull are the minimal but necessary set of particles that would lead to the decay of an extremal RN-AdS black hole. 

\section{Comments on the AdS WGC in Theories with Extended Supersymmetry}
\label{sec:commentsonBPSbound}

In Minkowski space, the strong version of WGC proposes that the WGC bound is saturated only in supersymmetric theory by BPS states (see e.g. \cite{Arkani-Hamed:2006emk,Harlow:2022ich}). 
The reason is that in theories with extended supersymmetry, the BPS bounds forbid strictly superextremal particles, so the only states that satisfy the WGC were precisely those saturating it, i.e., the BPS states. This suggests that extremal black holes are marginally stable only if they are BPS. Moreover, it has been shown that when the BPS bound and the extremality bound coincide, the tower WGC necessitates an infinite tower of BPS particles. \cite{Heidenreich:2015nta,Alim:2021vhs,Cota:2022maf,Gendler:2022ztv,FierroCota:2023bsp}. 

In Conjecture \ref{conj:WGCinAdS}, we found that the charge-to-mass ratio that a particle must satisfy in order to allow a RN-AdS black hole to decay is greater than the bound satisfied by the WGC in flat space. This would raise the question of how the AdS WGC we are proposing behaves in theories with extended supersymmetry, where all the states must satisfy a certain BPS bound.

In flat space, the only states that can satisfy the WGC and the BPS bound are those saturating both. This is different in AdS space, where we have shown that the black hole extremality condition no longer agrees with the black hole decay condition for charged particles. In fact, the requirement on the charge-to-mass ratio of the particle is stronger, which can be understood as a consequence of extremal AdS black holes having lower charge-to-mass ratios than in flat space, which makes the production of charged particles through the gauge field more difficult. The stronger WGC bound on particles in AdS raises the question of whether it is consistent with the BPS bound. In fact, it has been shown (e.g. in \cite{Romans:1991nq,Chamblin:1999tk,Hristov:2011ye}) that supersymmetric $\mathcal{N}=2$ supergravity solutions satisfy the same BPS condition as in flat space given by $\mc{Z}=1$, which means that these BPS solutions are superextremal -- they do not have any horizons and contain naked singularities. One would expect the naked singularity to be resolved when the full string theory solution with $\alpha'$ corrections is considered. However, the symmetry-protected relation between the charge and mass of the BPS solutions will not be corrected, which makes it seemingly impossible for a supersymmetric theory to satisfy the AdS WGC bound.

We will try to solve this puzzle, showing that the AdS WGC in Conjecture \ref{conj:WGCinAdS} allows BPS particles to be responsible for the decay of an extremal (non-BPS) RN-AdS black hole.\footnote{As pointed out in \cite{Lust:2019zwm}, scale-separated supersymmetric AdS vacua might not exist. This possibility will be discussed at the end of this section.} The subtlety lies in the notion of mass in AdS space. For large classical objects such as the SUGRA solutions, the mass can be identified with the ADM mass, leading to the BPS condition of $\mc{Z}=1$. For light states that have a Compton wavelength comparable to the AdS length, the energy should be associated with the conformal dimension $\Delta$ (i.e., the rest energy of a particle in AdS) rather than the mass parameter \cite{Witten:1998qj}. This leads to a BPS bound that is given in terms of $\Delta$ instead of $\mf{m}$, that in appropriate units reads  (see e.g. \cite{Aharony:1999ti,Berenstein:2002ke,Denef:2009tp,Hartnoll:2009sz,Nastase:2015wjb,Aharony:2021mpc}) 
\beq\label{eq:DeltaBPSboundwithq}
\Delta\gtrsim q\coma
\eeq
where $q$ is, usually, the R-charge. We expect the BPS bound to take the same form for a general $\U(1)$, which can satisfy the AdS WGC bound.
Recall the WGC bound
\beq \label{eq:WGCinBPSsection}
\frac{g_\UoD{D}^2\mf{q}^2M_\PlD{D}^{D-2}}{\gamma\mf{m}^2}\ge  \frac{1+\frac{(D-1)(D-2)}{(D-3)^2}\eta^2}{1+\frac{D-1}{D-3}\eta^2} \equiv \mathcal{R}(\eta)^2
\eeq
while, restoring the units in \eqref{eq:DeltaBPSboundwithq}, the AdS BPS bound reads\footnote{Note that this is precisely opposite to the earlier AdS WGC bound we reviewed in Section \ref{sec:WGCinAdS-previous}.} \cite{Denef:2009tp,Hartnoll:2009sz,Nakayama:2015hga}
\beq \label{eq:AdS_BPS}
\frac{g_\UoD{D}^2\mf{q}^2M_\PlD{D}^{D-2}}{\gamma \ell_\AdSD{D}^{-2}\Delta^2} \leq 1\fstop
\eeq
In this way, we can check under which conditions, states satisfying \eqref{eq:AdS_BPS}, can also satisfy Conjecture \ref{conj:WGCinAdS}. Recall that, as we wrote above, in flat space, the only states that satisfy both the WGC bound and the BPS bound are those that saturate the two. In AdS, the mass of a particle can be expressed in terms of $\Delta$ by using \eqref{eq:conformaldimension}, obtaining 
\beq\label{eq:masstoDelta}
\mf{m}^2\ell_\AdSD{D}^2=\Delta^2\left(1-(D-1)\Delta^{-1}\right)\coma
\eeq
giving the possibility of some $\Delta$ that would satisfy (or at least saturate) both the BPS bound and the WGC bound. In particular, for our purposes, it is sufficient to focus on BPS states, and check under what conditions on $\Delta$ they satisfy (or saturate) Conjecture \ref{conj:WGCinAdS}. Let us, then, write the WGC bound in \eqref{eq:WGCinBPSsection} as
\begin{equation}
    \frac{g_\UoD{D}^2\mf{q}^2M_\PlD{D}^{D-2}}{\gamma\ell_\AdSD{D}^{-2}\Delta^2}\frac{\ell_\AdSD{D}^{-2}\Delta^2}{\mf{m}^2}  \geq \frac{1+\frac{(D-1)(D-2)}{(D-3)^2}\eta^2}{1+\frac{D-1}{D-3}\eta^2}\equiv \mathcal{R}(\eta)^2\coma
\end{equation}
and focus on states that saturate \eqref{eq:AdS_BPS}, i.e. BPS states. In this case, we must require
\begin{equation}
    \frac{\ell_\AdSD{D}^{-2}\Delta^2}{\mf{m}^2} = \frac{1}{1+(1-D)\Delta^{-1}} \geq  \mathcal{R}(\eta)^2\coma
\end{equation}
where we have used the relation \eqref{eq:masstoDelta}. For a generic value of the horizon radius, the BPS states that satisfy the WGC are those whose $\Delta$ is 
\begin{equation}\label{eq:DeltaBoundWGC}
    D-1<\Delta \leq \frac{\mathcal{R}(\eta)^2(D-1)}{\mathcal{R}(\eta)^2-1} = (D-2) (D-1)+\frac{(D-3)^2}{\eta ^2}\fstop
\end{equation}
In particular, the WGC is saturated when
\begin{equation}
    \Delta = (D-2) (D-1)+\frac{(D-3)^2}{\eta ^2}\fstop
\end{equation}
So, a BPS state with conformal dimension $\Delta$ given by the expression above, will also saturate the WGC in AdS predicted in Conjecture \ref{conj:WGCinAdS}. Smaller conformal dimensions will be associated (if they exist) to particles that will satisfy the WGC. As $\Delta$ approaches the lower bound $D-1$, the corresponding mass $\mf{m}$ approaches zero (in AdS units).\footnote{Charged massless states trivially satisfy the WGC. In flat space, charged particles can become massless in special loci, e.g., at the intersection of Coulomb and Higgs branches in $\mathcal{N}=2$ theories.}
In fact, in terms of $\mf{m}$, the states that BPS states that satisfy Conjecture \ref{conj:WGCinAdS} have
\begin{equation}
\scalebox{0.93}{$\displaystyle
    0<\mf{m}^2\ell_\AdSD{D}^2\leq \frac{\mathcal{R}(\eta)^2(D-1)^2}{(\mathcal{R}(\eta)^2-1)^2} = (D-1)^2(D-2)  (D-3)+ \frac{(D-1) (2 D-5) (D-3)^2}{\eta ^2}+\frac{(D-3)^4}{\eta ^4}\fstop$}
\end{equation}

The strictest condition to satisfy is given in the limit $\eta \rightarrow \infty$. In that case, we have that
\begin{equation}\label{eq:Deltaallowedetainfty}
    D-1<\Delta \leq \lim_{\eta\rightarrow\infty}\frac{\mathcal{R}(\eta)^2(D-1)}{\mathcal{R}(\eta)^2-1} = (D-2) (D-1)\coma
\end{equation}
or analogously, the mass of the particle cannot be larger than
\begin{equation}
    \mf{m}^2\ell_\AdSD{D}^2 = (D-1)^2(D-2)(D-3)\fstop
\end{equation}

We have therefore found possible BPS states that satisfy the WGC we propose in Conjecture \ref{conj:WGCinAdS}. The heaviest possible BPS states (in AdS units) that satisfy the WGC and the BPS bounds are the ones that saturate both bounds, generalizing the result in flat space. Smaller conformal dimensions will be associated with states that have a smaller and smaller mass, asymptotically becoming massless when $\Delta \rightarrow D-1$. Thus, the BPS states satisfying the WGC are light and have a Compton wavelength comparable to the AdS scale. Therefore, the omitted piece in \eqref{eq:BF_with_O1} should be restored, leading to the requirement that
   \begin{equation}\label{eq:correctedWGCboundBPS}
        \frac{g_\UoD{D}^2\mf{q}^2M_\PlD{D}^{D-2}}{\gamma\ell_\AdSD{D}^{-2}\Delta^2}\frac{\ell_\AdSD{D}^{-2}\Delta^2}{\mf{m}^2}\geq \mathcal{R}(\eta)^2\left(1+\frac{1}{4\ell_\AdSD{2}^2\mf{m}^2}\right)\coma
    \end{equation}
where the $\AdS_2$ and the $\AdS_D$ length scales are related by \eqref{eq:AdS_length2d}, reading
\begin{equation}
    \ell_\AdSD{2} = \ell_\AdSD{D}\frac{\eta }{\sqrt{(D-3)^2+(D-2) (D-1) \eta ^2}} \equiv \frac{\ell_\AdSD{D}}{\mathcal{D}(\eta)}\coma
\end{equation}
where we have introduced $\mathcal{D}(\eta)$ for later convenience. We can repeat the computation we did before taking into account the correction coming from the BF bound, and asking ourselves what are the BPS states that also satisfy the WGC, namely 
\begin{equation}
    \frac{1}{1+(1-D)\Delta^{-1}}\geq \mathcal{R}(\eta)^2\left(1+\frac{\mathcal{D}(\eta)^2}{4\Delta^2\left(1+(1-D)\Delta^{-1}\right)}\right)\fstop
\end{equation}
The result is a smaller window of allowed conformal dimensions $\Delta$, given by
\begin{equation}
   D-1<  \Delta \leq \frac{\mathcal{R}(\eta)^2(D-1)+\sqrt{(\mathcal{R}(\eta)^4(D-1)^2- \mathcal{R}(\eta)^2\mathcal{D}(\eta)^2 (\mathcal{R}(\eta)^2-1)}}{2(\mathcal{R}(\eta)^2-1)}\fstop
\end{equation}
Clearly this condition reduces to \eqref{eq:DeltaBoundWGC} if one does not consider the correction coming from the BF bound. In the large black hole limit, $\eta\rightarrow \infty$, the bound becomes 
\begin{equation}\label{eq:Deltaboundwithcorrection}
    D-1 < \Delta \leq \frac{(D-2)(D-1)}{2} +\frac{(D-2)}{2}\sqrt{(D-2)(D-1)}\coma
\end{equation}
that shows a narrower window of allowed $\Delta$, compared to that obtained in \eqref{eq:Deltaallowedetainfty}. The WGC we propose is satisfied by BPS states whose conformal dimensions are in the range given by \eqref{eq:Deltaboundwithcorrection}, with decreasing mass of the states when $\Delta\rightarrow D-1$.

However, we leave open whether there exist supersymmetric theories that admit BPS states whose conformal dimensions are in this range of values. The result is based on our starting assumption that we are working with an EFT in AdS without any reference to the UV completion, meaning that it has been possible to integrate out all the massive modes of the internal space. In the case of supersymmetric EFTs in AdS, the AdS Distance Conjecture (ADC) \cite{Lust:2019zwm} states that there are no supersymmetric AdS vacua with separation of scales between the AdS length and the KK scale of the internal space.\footnote{This is in line with the recent findings for $\mathcal{N}=1$ supersymmetric AdS vacua \cite{Lust:2022lfc,Bena:2024are,Montero:2024qtz}.} This implies that our set-up of a single probe particle in a RN-AdS black hole background may be drastically modified by the presence of infinite towers of KK modes that are of the same scale of the AdS scale length. The expectation is that Conjecture \ref{conj:WGCinAdS} applies to EFTs with negative cosmological constant whenever possible to integrate out the KK towers of the internal space. 

\section{Discussion and Conclusions}
\label{sec:conclusions}

The main result of this paper is a version of the WGC for particles in AdS space. The result is summarized in Conjecture \ref{conj:WGCinAdS} and we now re-state it here:
\begin{mdframed}[backgroundcolor=white, shadow=true, shadowsize=4pt,shadowcolor=seccolor,
roundcorner=6pt]
    \begin{conjecture*}[AdS Weak Gravity Conjecture]
    Given any $\U(1)$ gauge field coupled to Einstein--Maxwell gravity, there must exist a particle of charge $\mf{q}$ and mass $\mf{m}$ such that
    \begin{equation}
       \frac{g_\UoD{D}\mf{q}}{\mf{m}}\ge \sqrt{\gamma}\frac{1}{M_\PlD{D}^{\frac{D-2}{2}}} \sqrt{\frac{1+\frac{(D-1)(D-2)}{(D-3)^2}\eta^2}{1+\frac{D-1}{D-3}\eta^2}}\coma \text{with } \eta = \frac{r_h}{\ell_\AdSD{D}}\coma
    \end{equation}
   where $r_h$ is the horizon of an extremal RN-AdS black hole and $\ell_\AdSD{D}$ is the $\AdS_D$ scale length. The value $\gamma$ is the extremality factor for a RN black hole in flat space, i.e. $\gamma  = \frac{D-3}{D-2}$. For particles with a mass comparable to the AdS scale, we refer to Remark \ref{remark:BFbound}.
Considering the largest possible RN-AdS black hole, i.e. $r_h\gg \ell_\AdSD{D}$, we require that there exists a particle of charge $\mf{q}$ and mass $\mf{m}$ such that 
    \begin{equation}
       \frac{g_\UoD{D}\mf{q}}{\mf{m}}\ge \frac{1}{M_\PlD{D}^{\frac{D-2}{2}}}\fstop
    \end{equation}
\end{conjecture*}
\end{mdframed}
\vspace{3mm}
 As we mentioned in the introduction, in flat space, if the gauge coupling depends on the moduli, the extremality factor changes. However, in this work, we focused on $\AdS_D$ Einstein--Maxwell theory, without any reference to the quantum gravity uplift from which it could have originated, and thus choosing to be agnostic on how the corrections to the extremality bound will enter in our expression for the AdS WGC. 

The way in which we argued for this conjecture is threefold\begin{enumerate*}[before=\unskip{: }, itemjoin={{. }}, label={({\arabic*})}, ref={{\arabic*}}]
    \item First, we extended the computation carried out in \cite{Lin:2024jug} to obtain the conditions for Schwinger pair production in an RN-AdS black hole background. This gave, for the first time, the condition on the charge-to-mass ratio that precisely enters into Conjecture \ref{conj:WGCinAdS}
    \item Then, we argued that the Schwinger effect is related to the instability of charged scalar fields that have an effective mass below the BF bound in the near-horizon $\AdS_2\times S^{D-2}$ geometry of the RN-AdS. Requiring the particles to violate the BF bound led to the same constraint on their charge-to-mass ratio
    \item Finally, we considered the conditions under which a probe particle experiences repulsion at the horizon of an extremal RN-AdS black hole, again confirming the WGC bound we obtained.
\end{enumerate*}

Furthermore, we discussed the extension of the conjecture in the case of multiple $\U(1)$, which led to the formulation of an AdS Convex Hull WGC as in Conjecture \ref{conj:CHCAdS}.
We also showed in Section \ref{sec:commentsonBPSbound} that BPS states in AdS are compatible with Conjecture \ref{conj:WGCinAdS}, and, as in flat space, they can also saturate the bound. However, our results are based on EFTs in which it has been possible to integrate out all the massive modes of the internal space, leading to an EFT in AdS. If one assumes the AdS Distance Conjecture (ADC) \cite{Lust:2019zwm}, our set-up of a single probe particle in a RN-AdS black hole background must be revisited because of the presence of the infinite towers of KK modes at the same scale of the AdS scale length. Our expectation is that Conjecture \ref{conj:WGCinAdS} applies to EFTs with negative cosmological constant whenever possible to integrate out the KK towers of the internal space.

\subsubsection*{Future Directions}

In this paper, we take the WGC as the requirement for extremal black holes to decay by emitting charged particles. Instead of light particles, the state satisfying the WGC can potentially be black holes because the extremality bound receives higher derivative corrections. This possibility has been extensively studied for extremal black holes in asymptotically flat spacetime \cite{Kats:2006xp, Hamada:2018dde, Loges:2020trf,Aalsma:2020duv, Cremonini:2020smy,Aalsma:2022knj}. 
However, as we have shown in this work, kinematic considerations alone do not determine whether black hole decay in AdS spacetime is possible. To derive the conditions for extremal black holes to decay into smaller black holes in AdS spacetime, one would have to identify the borderline case where two or more black holes experience no force. Such multi-centered black hole solutions in AdS spacetime (analog of the Majumdar--Papapetrou solution for Einstein--Maxwell theory in flat spacetime \cite{Majumdar:1947eu,Papaetrou:1947ib}) have not been found, though an interesting attempt has been made \cite{Cai:2022qac}. 

There exist many versions of the WGC in flat space. In this work, we showed that the extremal black hole decay formulation of WGC in AdS is equivalent to a RFC between an RN-AdS black hole and a particle. Our proposal of the AdS WGC aligns closely with the ideas in the original proposal of the WGC in flat space and its extension in the case of multiple $\U(1)$s. It is known that in the presence of massless scalar fields, the WGC and RFC differ, but including the contributions coming from the scalars, the towers of states satisfying the tower WGC in flat space are also those satisfying the RFC in flat space. It would be interesting to check how our conjecture is modified when there are scalar fields. In particular, when the scalar fields originate from dimensional reduction, this would require re-evaluating both the black hole background solution and the particle decay condition. In theory, one could follow the logic explained in \cite{Heidenreich:2015nta}, repeat the computations in our paper, and compactify the theory on a circle to check if Conjecture \ref{conj:CHCAdS} is violated. However, there are further subtleties when considering our bound in the context of circle compactification. For instance, the geometries of $\AdS_D$ and $\AdS_{D-1}\times S^1$ are not equivalent and one can make different choices of whether the circle dimension is warped or not. It will be interesting to understand how the geometry and black holes therein affect our proposal of the AdS WGC bound.

Regarding a tower of infinite particle states, one interesting consequence of Conjecture \ref{conj:CHCAdS} is that if the only particles in the spectrum are those saturating our proposed WGC bound for the decay of an extremal RN-AdS, then, in order for Conjecture \ref{conj:CHCAdS} to be satisfied, we would need infinitely many particles, so that all possible RN-AdS can decay. This statement is reminiscent of a tower version of the WGC in the presence of only BPS states. However, Conjecture \ref{conj:CHCAdS} is based on considering the decay of an extremal RN-AdS black hole by a particle, while the CHC in flat space \cite{Cheung:2014vva} considered the complete decay of black holes into a set of particles. This difference makes the motivation for the existence of an infinite tower of states satisfying the WGC different from the original motivation that led to the tower WGC in flat space. 

We reserve these interesting questions for future investigations.

\acknowledgments

The authors thank Nikolay Bobev, Craig Lawrie, Matteo Lotito, Jorge Santos, Chiara Toldo, Cumrun Vafa, Stefan Vandoren and Victoria Venken for helpful discussions and comments. This work is supported in part by the DOE grant DE-SC0017647.

\appendix

\bibliographystyle{JHEP}
\bibliography{references}

\end{document}